\documentclass[11pt,a4paper]{article}
\pdfoutput=1

\usepackage{jheppub}\usepackage{graphicx,placeins,amsmath,amssymb,subfigure,color,ulem}
\usepackage[applemac]{inputenc}
\usepackage{cancel}

\newcommand{\blue}[1]{\textcolor{blue}{#1}}

\begin{document}
\preprint{YITP-18-124, RIKEN-QHP-400, RIKEN-iTHEMS-Report-18}

\title{
Consistency between L\"uscher's finite volume method and HAL QCD method for two-baryon systems in 
lattice QCD
}

\author[a]{Takumi~Iritani,}
\author[b,c]{Sinya~Aoki,}
\author[a,d]{Takumi~Doi,}
\author[a,d]{Tetsuo~Hatsuda,}
\author[e]{Yoichi~Ikeda,}
\author[f]{Takashi~Inoue,}
\author[e]{Noriyoshi~Ishii,}
\author[e]{Hidekatsu~Nemura,}
\author[b]{and Kenji~Sasaki}
\author{(HAL QCD Collaboration)}

\affiliation[a]{ RIKEN Nishina Center, Wako 351-0198, Japan }
\affiliation[b]{ Center for Gravitational Physics, 
Yukawa Institute for Theoretical Physics, Kyoto University, Kitashirakawa Oiwakecho, Sakyo-ku, 
Kyoto 606-8502, Japan }
\affiliation[c]{ Center for Computational Sciences, University of Tsukuba, Tsukuba 305-8577, Japan }
\affiliation[d]{ RIKEN Interdisciplinary Theoretical and Mathematical Science Program (iTHEMS), Wako 351-0198, Japan }
\affiliation[e]{ Research Center for Nuclear Physics (RCNP), Osaka University, Osaka 567-0047, Japan }
\affiliation[f]{ Nihon University, College of Bioresource Sciences, Kanagawa 252-0880, Japan }
\emailAdd{takumi.iritani@riken.jp}
\emailAdd{saoki@yukawa.kyoto-u.ac.jp}
\emailAdd{doi@ribf.riken.jp}
\emailAdd{thatsuda@riken.jp}
\emailAdd{yikeda@rcnp.osaka-u.ac.jp}
\emailAdd{inoue.takashi@nihon-u.ac.jp}
\emailAdd{ishiin@rcnp.osaka-u.ac.jp}
\emailAdd{hidekatsu.nemura@rcnp.osaka-u.ac.jp}
\emailAdd{kenjis@yukawa.kyoto-u.ac.jp}

\abstract{
 There exist two methods to study two-baryon systems in lattice QCD:
the   direct method which extracts eigenenergies from the plateaux of the temporal correlation function
 and the HAL QCD method which extracts observables from the non-local potential associated  with the 
 tempo-spatial correlation  function.  Although the two methods should  give the same results theoretically,
 there have been reported qualitative difference for observables from lattice QCD simulations.
 Recently, we pointed out in~\cite{Iritani:2016jie, Iritani:2017rlk} that
  the separation of the ground state from the  excited states is crucial to obtain sensible results in the former, 
  while both states provide useful signals for observables in the latter.
 In this paper, we  identify  the contribution of  each  state  in the direct method by decomposing
  the two-baryon correlation functions into  the finite-volume eigenmodes  obtained from the HAL QCD method.
   As in our previous studies, we consider  the $\Xi\Xi$ system in the $^1$S$_0$ channel at $m_\pi = 0.51$~GeV
  in (2+1)-flavor lattice QCD using the wall and smeared quark sources with spatial extents, $La = 3.6, 4.3, 5.8$~fm.    
  We demonstrate that   the ``pseudo-plateau'' at early time slices ($t = 1 \sim 2$~fm) from the smeared source in the direct method
   indeed originates  from  the contamination of the  excited states, and 
   the plateau with the ground state saturation is realized only at 
  $t > 5 \sim 15$~fm corresponding to the inverse of the lowest excitation energy.
  We also demonstrate that the two-baryon operator can be optimized by 
  utilizing the finite-volume eigenmodes, so that
 (i) the finite-volume energy spectra from the HAL QCD method agree with those from the temporal correlation function
 with the optimized operators and (ii) the correct finite-volume spectra  would be accessed in the direct method only if
    highly optimized operators are employed.
 Thus we conclude that the long-standing issue on the consistency
 between L\"uscher's finite volume method and the HAL QCD method for two baryons is now resolved
  at least for this particular system considered here:
  They are consistent with each other quantitatively  only if the excited contamination is properly removed in the former.
}


\maketitle

\section{Introduction}
\label{sec:introduction}

The interactions between two baryons have been studied
by two methods in lattice QCD.
The first one is the direct method~\cite{Yamazaki:2015asa,Wagman:2017tmp,Berkowitz:2015eaa},
which extracts the eigenenergies of the ground and/or the excited states
from the temporal correlations of two-baryon systems.
The binding energies and scattering phase shifts are calculated
from eigenenergies using L\"uscher's finite volume formula~\cite{Luscher:1985dn,Luscher:1990ux}.
The second one is the HAL QCD method~\cite{Ishii:2006ec,Aoki:2009ji,HALQCD:2012aa,Aoki:2012tk},
which derives the energy-independent non-local  kernel  (called the ``potential" in the literature) 
from the tempo-spatial correlations of two baryons.
Then the binding energies and phase shifts in the infinite volume are calculated by using 
the Schr\"{o}dinger-type equation with the kernel as the potential, 
which has field theoretical derivation on the basis of the reduction formula for composite operators.
Both methods rely on the asymptotic behavior of the Nambu-Bethe-Salpeter (NBS) wave function,
and should  in principle give the same results for observables~\blue{\cite{Aoki:2009ji,Aoki:2012tk,Aoki:2010ry}}.
In practice, however, the current numerical results for two-nucleon ($NN$) systems
seem to be inconsistent  with each other:
 For heavy pion masses ($m_\pi > 0.3$~GeV), 
both dineutron ($^1$S$_0)$ and deuteron ($^3$S$_1$)
are claimed to be bound in the direct method,
while those are unbound in the HAL QCD method.
 Also, the discrepancy is ubiquitous in two-baryon systems:
 Although both methods indicate a bound H-dibaryon in the SU(3) flavor limit
at $m_\pi = m_K \simeq 0.8$~GeV,
the binding energy is
74.6(4.7)~MeV
in the direct method~\cite{Beane:2012vq}, while
 it is 37.8(5.2)~MeV
in the HAL QCD method~\cite{Inoue:2011ai}.\footnote{Recently, 
  there appears another study~\cite{Francis:2018qch}
using the direct method, which indicates 
the dineutron is unbound
while the H-dibaryon is bound 19(10)~MeV
 at $m_\pi = m_K = 0.96$~GeV.
}

In a series of recent papers~\cite{Iritani:2016xmx,Iritani:2016jie,
  Iritani:2017rlk,Iritani:2017wvu,Aoki:2017byw,Iritani:2018zbt},
we have carefully examined the systematic uncertainties in both methods.
The difficulty of two-baryon systems
compared to a single baryon
originates from
the existence of elastic scattering states.
Their typical excitation energies $\delta E$ are one to two orders of magnitude
smaller than ${\cal O}(\Lambda_{\rm QCD})$, so that
 one needs to probe  large   Euclidean time $t \gtrsim (\delta E)^{-1}$ to extract the genuine signal of the 
 ground state in the direct method.
  However, the statistical fluctuation increases exponentially  in  $t$ as well as 
  the baryon number $A$ for multi-baryon systems as proved in ~\cite{Parisi:1983ae, Lepage:1989hd}.
  This practically prevents one to identify the signal of the ground state in the naive analysis of the temporal correlation of two baryons.

Moreover, our extensive studies~\cite{Iritani:2016jie, Iritani:2017rlk}
showed that a commonly employed procedure in the direct method to identify plateaux  at early time slices,  $t \ll (\delta E)^{-1}$, 
suffers from uncontrolled systematic errors from the excited state contaminations,
since pseudo-plateaux
\footnote{
    In Refs.~\cite{Iritani:2016jie,Iritani:2017rlk}, they are called ``fake plateaux'' or
    ``mirages'' of the plateau of the ground state.
}
easily appear at early time slices.
The typical symptoms of such systematics in the previous studies 
were explicitly exposed by the normality check%
\footnote{
    In Ref.~\cite{Iritani:2017rlk}, it is called ``sanity check'', a common terminology in computer science
    for a simple/quick test~\cite{wiki:sanity}.
}
based on L\"uscher's finite volume formula~\cite{Luscher:1990ux}
and the analytic properties of the $S$-matrix~\cite{Iritani:2017rlk}.

As far as the HAL QCD method is concerned,
the time-dependent formalism~\cite{HALQCD:2012aa}
is free from the problem of the ground state saturation,
since
the energy-independent potential
is extracted from the spatial and temporal correlations
with the information of both the ground and excited states associated with the elastic scattering.
While in practical calculations
there appears a systematic uncertainty associated with the
truncation of the derivative expansion for the non-locality of the potential,
the derivative expansion is found to be well converged at low energies~\cite{Iritani:2018zbt,Murano:2011nz}.
Other systematic uncertainties such as the contaminations from the inelastic states
and the finite volume effect for the potential are also shown
to be well under control~\cite{Iritani:2018zbt}.

In this paper,
we reveal the
 origin of the inconsistency
between the direct method and the HAL QCD method,
by explicitly evaluating the magnitude of the excited states in the 
temporal correlation functions.
We focus on the $\Xi\Xi$ system in the $^1$S$_0$ channel, 
which is a most convenient channel to obtain insights into the $NN$ systems,
since it belongs to the same SU(3) flavor multiplet as $NN (^1S_0)$
but has much better statistical signals.
Detailed studies in this channel were already performed with the direct method~\cite{Iritani:2016jie} as well as
the HAL QCD method~\cite{Iritani:2018zbt} in (2+1) flavor lattice QCD
at $m_\pi = 0.51$~GeV and $m_K = 0.62$~GeV, so that 
the main purpose of this paper is to present an in-depth analysis by combining both results:
In particular,  the excited state contaminations in the temporal correlation functions
are quantitatively evaluated by decomposing them in terms of
the finite-volume eigenmodes of  Hamiltonian with the HAL QCD potential (the HAL QCD Hamiltonian).

We show how the pseudo-plateau actually appears at early time slices
and also predict the time slice  at which  the ground state saturation is achieved.
Moreover, we establish a consistency between the direct method and the HAL QCD method,   
by demonstrating that temporal correlation functions constructed from the optimized two-baryon operators by the eigenmode of the HAL QCD Hamiltonian show 
the plateaux with the values consistent with the eigenenergies at early time slices.

This paper is organized as follows.
In Sec.~\ref{sec:formalism},
we introduce the theoretical framework of the direct method and the HAL QCD method,
and present the numerical setup of the lattice QCD calculation.
In Sec.~\ref{sec:previous},
we recapitulate the previous analysis on the direct method~\cite{Iritani:2016jie}
 as well as on  the HAL QCD method~\cite{Iritani:2018zbt}.
In Sec.~\ref{sec:anatomy},
we decompose the correlation functions into the eigenmodes of 
the HAL QCD Hamiltonian.
The anatomy of
the excited state contaminations in the direct method is presented. 
We also demonstrate that eigenfunctions can be used to optimize two baryon-operators.
The consistency between the temporal correlations
with the optimized operators and 
 the HAL QCD method is established.
Sec.~\ref{sec:summary} is devoted to the conclusion.
In Appendix~\ref{app:n2lo},
we check how the 
 next-to-next-leading order (N$^2$LO) 
analysis for the HAL QCD potential affects the finite volume spectra.
In Appendix~\ref{app:eigen_func},
we collect  eigenfunctions of the HAL QCD Hamiltonian on various volumes.
In Appendix~\ref{app:inelastic},
we study the reconstruction of the $R$-correlator from the elastic states.
In Appendix~\ref{app:delta_E_r},
we collect the results for the reconstruction of the effective energy shifts.
In Appendix~\ref{app:eigen-proj},
we show the effective energy shifts from the optimized operators on various volumes.
We note that a preliminary account of this study was reported
in Refs.~\cite{Iritani:2016xmx, Iritani:2017wvu}.

\section{Methods and Lattice setup}
\label{sec:formalism}

In this section, we briefly summarize the direct method and the HAL QCD method
for  two-baryon systems, together with the lattice setup used in this paper.

\subsection{Direct method}
\label{subsec:formalism:direct}
In the direct method for two-baryon systems,
the energy eigenvalues (on a finite volume) are measured from the temporal correlation
of the two-baryon operator, $\mathcal{J}^{\rm sink, src}_{BB}(t)$, as
\begin{equation}
  C_\mathrm{BB}(t) \equiv \langle 0 | \mathcal{J}^{\rm sink}_{BB}(t) \overline{\mathcal{J}}^{\rm src}_{BB}(0)| 0 \rangle
  = \sum_n Z_n e^{-W_n t} + \cdots,
\label{eq:4pt_direct}
\end{equation}
where $W_n$ is the energy of $n$-th two-baryon elastic state
and the ellipsis denotes the inelastic contributions.
In order to obtain the energy shifts
$\Delta E_n \equiv W_n - 2m_B$
with $m_B$ being the single baryon mass,
one often uses the ratio of the temporal correlation function of two- (one-) baryon system
$C_\mathrm{BB}(t)$ ($C_\mathrm{B}(t)$) as 
\begin{equation}
  R(t) \equiv \frac{C_\mathrm{BB}(t)}{\{C_\mathrm{B}(t)\}^2}, \qquad
  C_B(t) = Z_B e^{-m_B t} + \cdots ,
\end{equation}
to reduce the statistical uncertainties as well as some systematics thanks to
the correlations between $C_\mathrm{BB}(t)$ and $C_\mathrm{B}(t)$.
The energy shift of the ground state can be obtained from
the plateau value of the effective energy shift defined by
\begin{equation}
  \Delta E_\mathrm{eff}(t) \equiv \frac{1}{a} \log \frac{R(t)}{R(t+a)} 
  \label{eq:Eeff}
\end{equation}
with $a$ being the lattice spacing.
Here $t$ needs to be  sufficiently large compared to the inverse of the excitation energy.

Once the energy shift of the ground (or excited) state
on a finite volume
is obtained,
one can calculate the scattering phase shift in the infinite volume
at that energy, $\delta_0 (k)$,
via L\"uscher's finite volume formula~\cite{Luscher:1990ux},
\begin{equation}
  k \cot \delta_0 (k) = \frac{1}{\pi (La)} \sum_{\vec n\in \mathbf{Z}^3}\frac{1}{\vec n^2 -q^2},
  \qquad q=\frac{k (La)}{2\pi}, 
  \label{eq:kcot_delta}
\end{equation}
where we  consider the S-wave scattering
for simplicity, 
$k$ is defined through
$W_n = 2\sqrt{m_B^2 + k^2}$
and $L$ is the number of the spatial site of the lattice box.
If the energy shift $\Delta E_n$ is negative,
the analytic continuation of the above formula to $k^2 <0$ is understood.
The state with a negative energy shift
in the infinite volume limit corresponds to a bound state.

As noted before,
the origin of the difficulty of two-baryon systems
is the existence of elastic scattering states.
Since the typical excitation energy of such states is $(2\pi)^2/((La)^2m_B)$,
the ground state saturation requires extremely large
$t$, e.g.,
$t \gtrsim {\cal O}(4)$~fm at $La=4$ fm and $m_B=2$~GeV,
where the bad signal-to-noise ratio makes it
practically impossible to obtain signals.
In the literature of the direct method~\cite{Yamazaki:2015asa, Wagman:2017tmp, Berkowitz:2015eaa}, however, one extracted the  energy shift for the ground state from 
the plateau-like behavior of the effective energy shift at  early time slices, $t \sim  {\cal O}(1)$~fm instead, 
assuming that the ground state saturation is achieved.

\subsection{HAL QCD method}
\label{subsec:formalism:hal}
In the HAL QCD method, 
the energy-independent non-local potential $U(\vec{r}, \vec{r'})$ is defined from
\begin{equation}
  (E_k - H_0) \psi^W(\vec{r})
  =\int d\vec{r'} U(\vec{r}, \vec{r'})
  \psi^W(\vec{r'}) ,
 \label{eq:SCH}
\end{equation}
with the Nambu-Bethe-Salpeter (NBS) wave function~\cite{Ishii:2006ec,Aoki:2009ji},
\begin{eqnarray}
\psi^W(\vec{r}) &=& \langle 0 \vert T\{ \sum_{\vec{x}} B(\vec{x}+\vec{r},0)B(\vec{x},0) \} \vert 2B, W \rangle.
\end{eqnarray}
Here $\vert 2B, W\rangle$ is the QCD eigenstate for two baryons
with the eigenenergy $W = 2\sqrt{m_B^2 + k^2}$ in the center of mass system,
$B(\vec{x},t)$ is a single baryon operator, 
$E_k = k^2/(2\mu)$, and $H_0 = -\nabla^2/(2\mu)$ with $\mu = m_B/2$ being the reduced mass.
Eq.~(\ref{eq:SCH}) has field theoretical derivation on the basis of the Nishijima-Zimmermann-Haag 
reduction formula for composite operators~\cite{Haag:1958vt}.
Below the inelastic threshold $W_\mathrm{th}$,
the potential $U(\vec{r}, \vec{r'})$ is shown to be faithful to the phase shifts,
which are encoded in the behaviors of the NBS wave functions at large~$r$.

The four-point correlation function of the two-baryon system $F(\vec{r}, t)$ is  given by
\begin{eqnarray}
  F(\vec{r}, t) &\equiv& \langle 0 | T \{ \sum_{\vec{x}} B(\vec{x}+\vec{r},t)B(\vec{x},t) \overline{\mathcal{J}}^{\rm src}_{BB}(0)\}| 0\rangle \\
  &=& \langle 0 | T\{ \sum_{\vec{x}} B(\vec{x}+\vec{r},t) B(\vec{x},t)\}
  \sum_{n} | 2B, W_n \rangle \langle 2B, W_n |
  \overline{\mathcal{J}}^{\rm src}_{BB}(0) | 0 \rangle + \cdots \nonumber \\
  &=& \sum_{n} A_{n} \psi^{W_n}(\vec{r})e^{-W_n t} + \cdots, 
\end{eqnarray}
where
$A_{n} \equiv \langle 2B, W_n|\overline{\mathcal{J}}^{\rm src}_{BB}(0)|0\rangle$
is the overlap factor
and the ellipsis represents the inelastic contributions.

In the time-dependent HAL QCD method~\cite{HALQCD:2012aa,Aoki:2012tk},
the potential is extracted directly from the so-called $R$-correlator as
\begin{equation}
  \left[ 
    -H_0 - \frac{\partial}{\partial t}
    + \frac{1}{4m_B} \frac{\partial^2}{\partial t^2}
  \right]R(\vec{r},t)
  = \int d\vec{r'} U(\vec{r}, \vec{r'}) R(\vec{r'},t),
  \label{eq:pot-def}
\end{equation}
where
\begin{equation}
  R(\vec{r}, t) \equiv \frac{F(\vec{r}, t)}{\{C_B(t) \}^2}
  =  \sum_{n} \frac{A_{n}}{Z_B^2} \psi^{W_n}(\vec{r})e^{-(W_n-2m_B) t} + \cdots ,
\end{equation}
 with the ellipsis being the inelastic contributions.
 Eq.~(\ref{eq:pot-def}) requires 
neither the ground state saturation nor  
the determination of individual  eigenenergy $W_n$ and eigenfunction $\psi^{W_n}(\vec{r})$,
as all elastic states can be used to extract the energy-independent potential.
Therefore, compared with the direct method, the condition required  for the reliable calculation  is much more relaxed  in the time-dependent HAL QCD method
as   $t$ $\gtrsim {\cal O}(\Lambda_{\rm QCD}^{-1}) \sim \mathcal{O}(1)$~fm,
 where 
$R(\vec{r},t)$ is saturated by the contributions from elastic states (``the elastic state saturation'').

In practice, we expand  the non-local potential in terms of derivatives as  $U(\vec{r}, \vec{r'}) = \displaystyle \sum_n V_n (\vec{r}) \nabla^n \delta(\vec{r} - \vec{r'})$.
The leading order (LO) approximation gives 
$
  U(\vec{r},\vec{r'}) \simeq V_0^\mathrm{LO}(r) \delta(\vec{r} - \vec{r'}),
$
which can be determined as 
\begin{equation}
  V_0^\mathrm{LO}(\vec{r}) =
  - \frac{H_0 R(\vec{r},t)}{R(\vec{r},t)}
  - \frac{(\partial/\partial t) R(\vec{r},t)}{R(\vec{r},t)}
  + \frac{1}{4m_B} \frac{(\partial/\partial t)^2 R(\vec{r},t)}{R(\vec{r},t)} .
  \label{eq:veff}
\end{equation}
We can also examine the effect of higher order contributions to observables such as 
the scattering phase shifts:
In this paper, 
we present the study on
the correction to the LO potential for the spin-singlet channel  at 
the next-to-next-leading order (N$^2$LO) as
\begin{eqnarray}
  U(\vec{r}, \vec{r'}) &\simeq& \{V_0^\mathrm{N^2LO}(\vec{r}) + V_2^\mathrm{N^2LO}(\vec{r})\nabla^2\}\delta(\vec{r}- \vec{r'}) .
  \label{eq:pot:N2LO}
\end{eqnarray}

\subsection{Lattice setup}
\label{sec:setup}
Numerical data in previous literature~\cite{Iritani:2016jie,Iritani:2018zbt} and in this paper are obtained 
from the (2+1)-flavor lattice QCD ensembles,
generated in Ref.~\cite{Yamazaki:2012hi}
with the Iwasaki gauge action
and nonperturbatively $\mathcal{O}(a)$-improved Wilson quark action
at the lattice spacing $a=0.08995(40)$~fm ($a^{-1} = 2.194(10)$~GeV).
 In the present  paper, we make use of gauge ensembles on three lattice volumes,
$L^3 \times T =$ $40^3 \times 48$, $48^3 \times 48$, and $64^3 \times 64$,
with 
heavy up and down quark masses and the physical strange quark masses,
corresponding to
$m_\pi = 0.51$~GeV,
$m_K = 0.62$~GeV,
$m_N = 1.32$~GeV and $m_\Xi = 1.46$~GeV.

We employ  two different quark sources with the Coulomb gauge fixing,
the wall source,
$q^\mathrm{wall}(t) = \sum_{\vec{y}}q(\vec{y},t)$,
mainly used in  the HAL QCD method,
and the smeared source,
$q^\mathrm{smear}(\vec{x},t) = \sum_{\vec{y}} f(|\vec{x}-\vec{y}|)q(\vec{y},t)$,
 often used in the direct method.
For the smearing function, we take $f(r) \equiv \{Ae^{-Br}, 1, 0\}$ for $\{0 < r < (L-1)/2$, $r=0$, $(L-1)/2\leq r\}$, respectively, as in Ref.~\cite{Yamazaki:2012hi},
and the center of the smeared source is same for all  six quarks
(i.e., zero displacement between two baryons),
as has been employed in all previous studies in the direct method
claiming the existence of the $NN$ bound states
for heavy quark masses~\cite{Yamazaki:2015asa,Wagman:2017tmp,Berkowitz:2015eaa}.%
\footnote{
  Ref.~\cite{Berkowitz:2015eaa} uses a non-zero displaced operator as well.
}
For both sources,
the point-sink operator for each baryon is employed in this study.
A number of configurations and other parameters are summarized in Table~\ref{tab:lattice_setup}.
The correlation functions are calculated by the unified contraction algorithm (UCA)~\cite{Doi:2012xd}
and the statistical errors are evaluated by the jack-knife method.
For more details on the simulation setup, see Ref.~\cite{Iritani:2016jie}.

In this paper, we focus on $\Xi\Xi$ ($^1$S$_0$) system, which
belongs to the same $\mathbf{27}$ representation as $NN$ ($^1$S$_0$) in the flavor SU(3)
 transformation,
but has much better signal-to-noise ratio than $NN$ as the system contains four strange quarks. 
We use the relativistic interpolating operator~\cite{Iritani:2016jie} for $\Xi$, given by
\begin{equation}
  \Xi_\alpha^0 = \varepsilon_{abc}(s^{aT}C\gamma_5 u^b)s_\alpha^c, \quad
  \Xi_\alpha^- = \varepsilon_{abc}(s^{aT}C\gamma_5 d^b)s_\alpha^c,
  \label{eq:Xi_op}
\end{equation}
where $C = \gamma_4\gamma_2$ is the charge conjugation matrix, 
$\alpha$ and $a$, $b$, $c$ are the spinor and color indices, respectively.

\begin{table}
  \centering
  \begin{tabular}{c|c|c|cc|c}
    \hline
    \hline
    volume & $La$ & \# of conf. & \# of smeared sources & $(A,B)$ & \# of wall sources \\
    \hline
    $40^3 \times 48$ & 3.6 fm & 207 & 512 & (0.8, 0.22) & 48 \\ 
    $48^3 \times 48$ & 4.3 fm & 200 & $4 \times 384$  & (0.8, 0.23) & $4 \times 48$ \\ 
    $64^3 \times 64$ & 5.8 fm & 327 & $1 \times 256$  & (0.8, 0.23) & $4 \times 64$ \\ 
    \hline
    \hline
  \end{tabular}
  \caption{Simulation parameters.
  The rotational symmetry for isotropic lattice is used to increase statistics.}
 \label{tab:lattice_setup}
\end{table}

\section{Summary of previous studies}
\label{sec:previous}

\subsection{Operator dependence of the plateaux in the direct method}
\label{subsec:direct}
\begin{figure}[hbt]
  \centering
  \includegraphics[width=0.47\textwidth,clip]{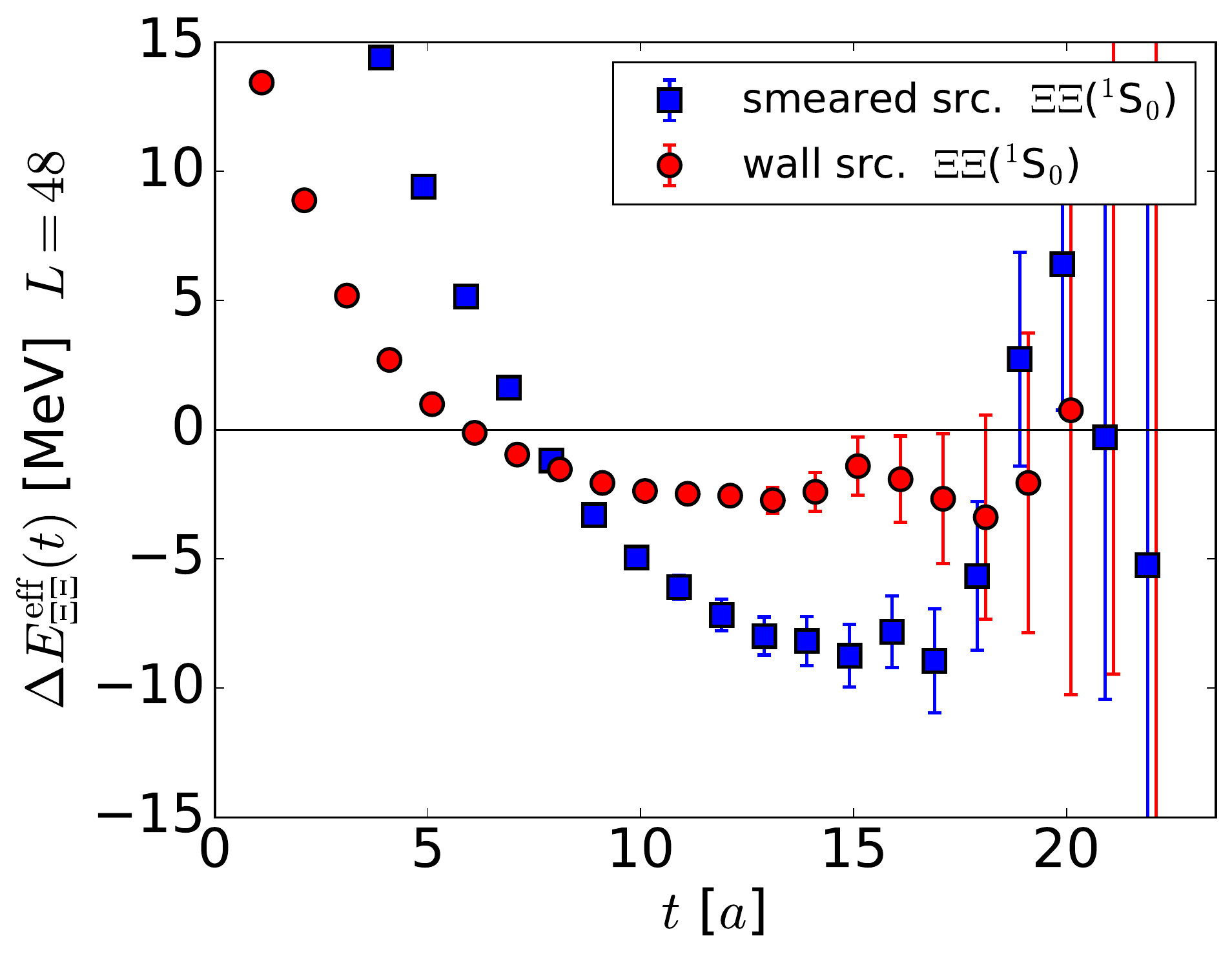}
   \includegraphics[width=0.47\textwidth,clip]{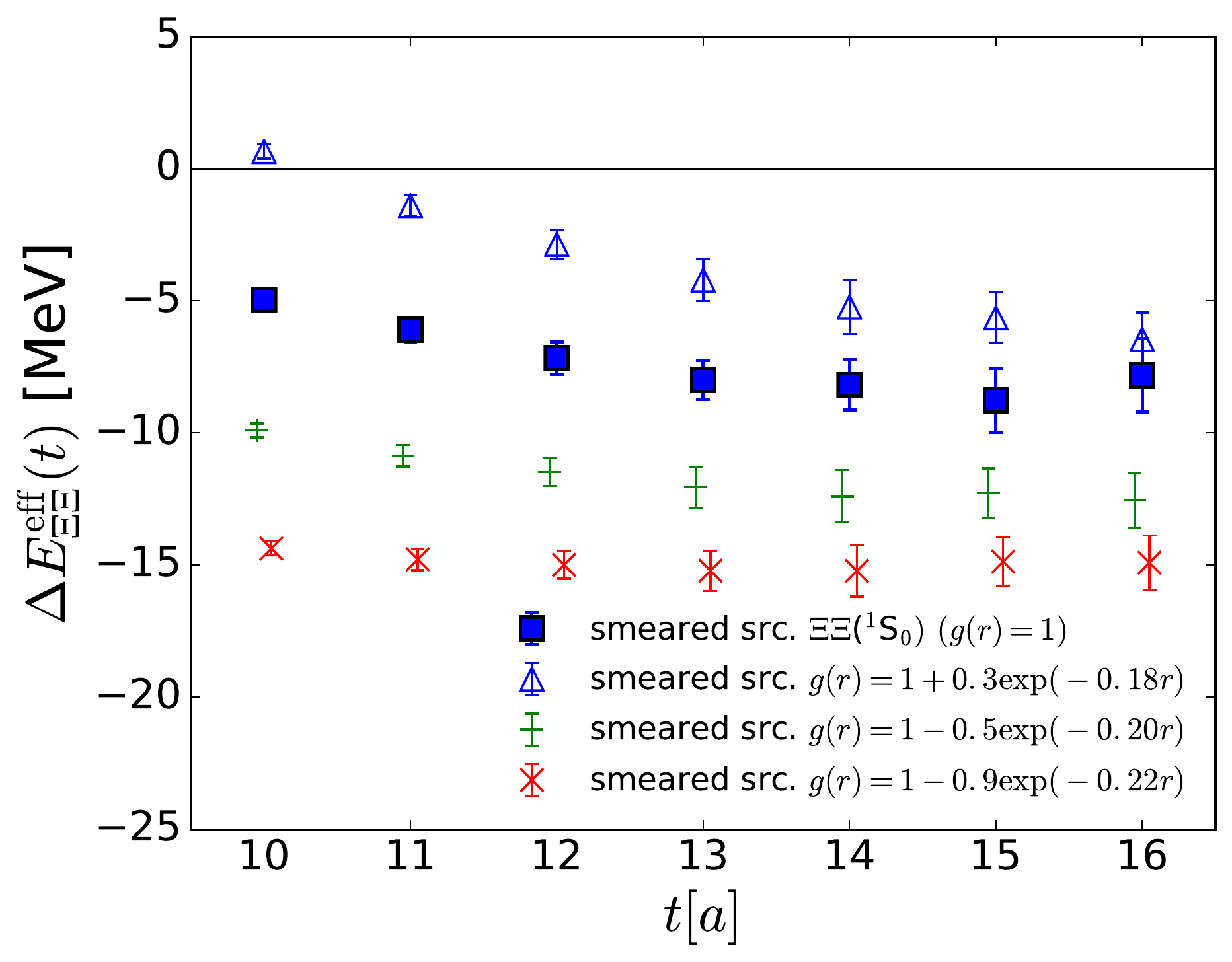}
  \caption{
    \label{fig:deltaEeff:src-dep}
    (Left) The source operator dependence of the effective energy shift $\Delta E_\mathrm{eff}(t)$
    for $\Xi\Xi$ ($^1$S$_0$) using the wall source (red circles) and the smeared source (blue squares) for $L=48$~\cite{Iritani:2016jie}.
    (Right) The sink operator dependence of the same quantity  with the smeared source~\cite{Iritani:2016jie}.
  }
\end{figure}

In Ref.~\cite{Iritani:2016jie}, we pointed out that the plateau-like behaviors at $t\simeq 1$ fm in the direct method depend on sources or sink operators.
For example, Fig.~\ref{fig:deltaEeff:src-dep} (Left) shows
the source operator dependence of
the effective energy shift $\Delta E_\mathrm{eff}(t)$ in Eq.~(\ref{eq:Eeff})
for the $\Xi\Xi$ ($^1$S$_0$) on $L=48$, where $ \mathcal{J}^{\rm sink}_{BB}(t)$ in Eq.~(\ref{eq:4pt_direct}) is given by
\begin{eqnarray}
 \mathcal{J}^{\rm sink}_{BB}(t) &=& \sum_{\vec{r}} \sum_{\vec{x}} B(\vec{x}+\vec{r},t)B(\vec{x},t) ,
 \label{eq:point_sink}
\end{eqnarray}
where a baryon operator $B$ is given in Eq.~(\ref{eq:Xi_op}).
While plateau-like structures appear around $t/a \sim 15$ for both wall and smeared sources,
the values disagree with each other.
Similar inconsistencies are found on other volumes
and the infinite volume extrapolation implies that
the system is bound (unbound) for the smeared (wall) source.
These discrepancies indicate some uncontrolled systematic errors.
Indeed, such an early-time pseudo-plateau can be shown to appear even with 
 10\% contamination of the excited states as demonstrated by 
 the mockup data~\cite{Iritani:2016jie}. 

Fig.~\ref{fig:deltaEeff:src-dep} (Right) shows the sink operator dependence of $\Delta E_\mathrm{eff}(t)$ for $\Xi\Xi$ ($^1$S$_0$) 
with the smeared source fixed,
where the sink operator is generalized as
\begin{eqnarray}
  \mathcal{J}^{\rm sink}_{BB}(t) &=& \sum_{\vec{r}} g(\vec{r}) \sum_{\vec{x}} B(\vec{x}+\vec{r},t)B(\vec{x},t) ,
  \label{eq:gen_op}
  \\
  g(\vec{r}) &=& 1 + \tilde{A} \exp(- \tilde{B} r) ,
\end{eqnarray}
with four different parameter sets,
$(\tilde{A},\tilde{B}) = (0.3,0.18), (-0.5,0.20), (-0.9,0.22)$ and $(0,0)$.
The last one corresponds to the simple sink operator ($g(r)=1$) in Eq.~(\ref{eq:point_sink}).
Although a plateau-like structure  is observed 
for each sink operator,
the values disagree with each other. This implies that the plateau-like behaviors at $t\simeq 1$ fm 
 with the smeared source
 are not the plateau of the ground state but are pseudo-plateaux
caused by contaminations of elastic scattering states other than the ground state.
We note that such sink-operator dependence is not observed  in the case of the wall source~\cite{Iritani:2016jie}.

\subsection{Normality check}
\label{subsec:sanity}
\begin{figure}[thb]
  \centering
  \includegraphics[width=0.47\textwidth,clip]{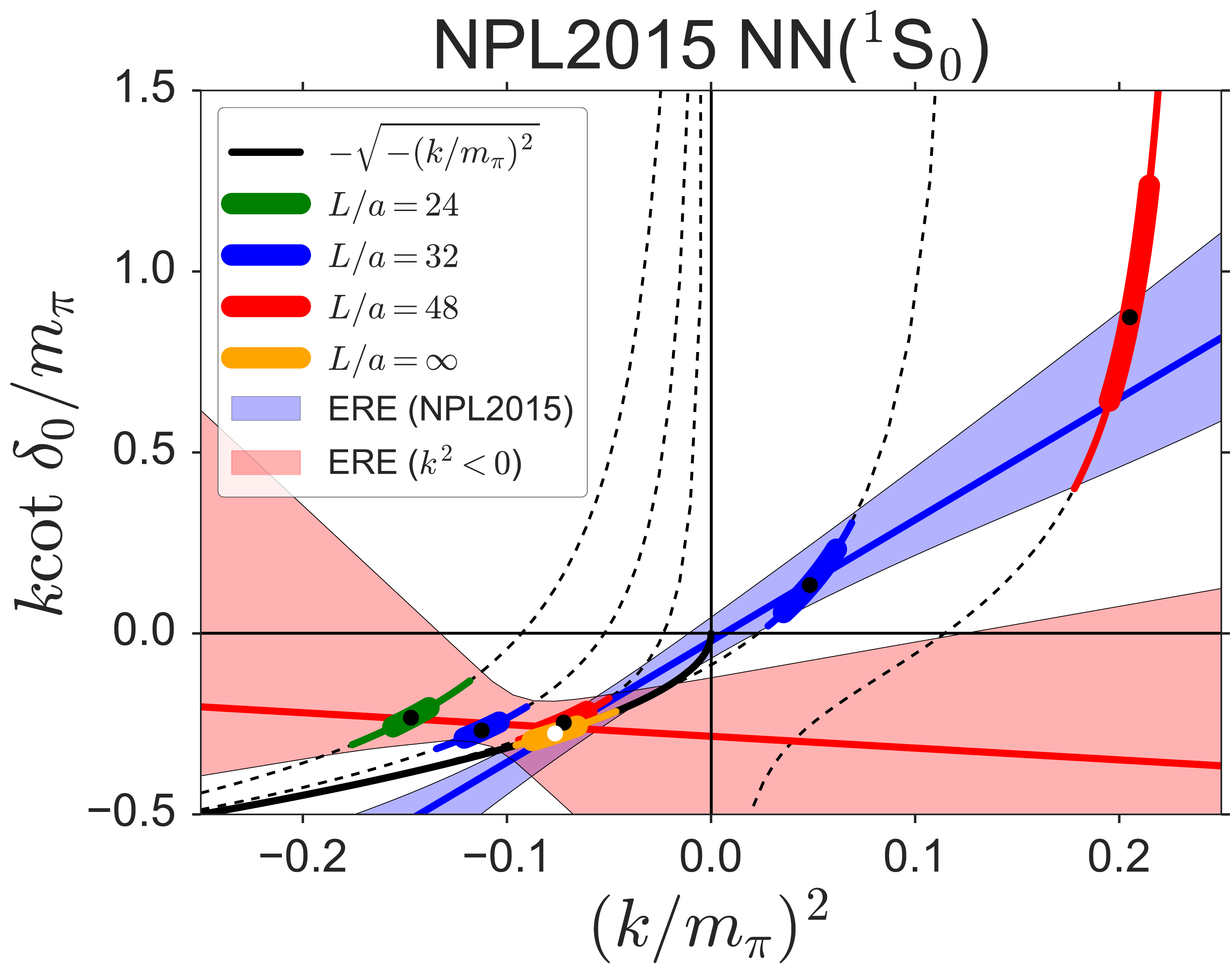}
  \includegraphics[width=0.47\textwidth,clip]{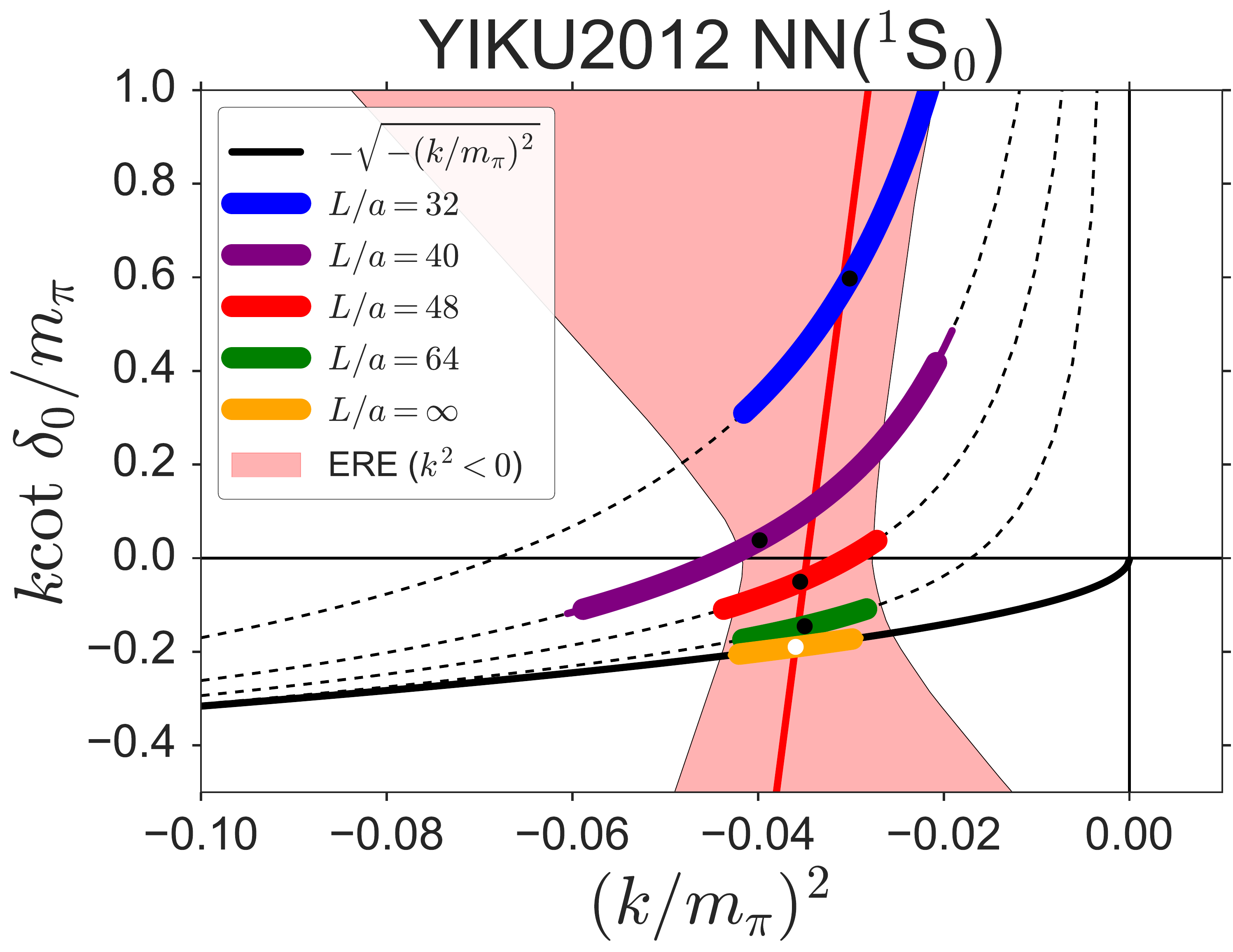}
  \caption{
    \label{fig:sanity}
    $k\cot\delta_0(k)/m_\pi$ as a function of $(k/m_\pi)^2$ for $NN$($^1$S$_0$)
     on each volume and the infinite volume in the direct method
    from Ref.~\cite{Orginos:2015aya} (Left) and Ref.~\cite{Yamazaki:2012hi} (Right).
    Black dashed lines correspond to L\"uscher's formula for each volume,
    while the black solid line represents the bound-state condition, $-\sqrt{-(k/m_\pi)^2}$.
    The red line (with an error band) corresponds to the ERE obtained from the data at $k^2 <0$ on finite volumes.
    In the left figure,
    the ERE fit to the data at $k^2 >0$ on finite volumes together with only  the infinite volume limit at $k^2<0$
    is also shown by the blue line.
    Both figures from Ref.~\cite{Iritani:2017rlk}.
  }
\end{figure}

Since the information on operator dependence as shown in the previous subsection is not always available,
we have introduced  the  ``normality  check''
in Ref.~\cite{Iritani:2017rlk},
based on L\"uscher's finite volume formula together with the analytic properties of the $S$-matrix. 

 Some examples of the normality check are given in Fig.~\ref{fig:sanity}, 
where $k\cot\delta_0(k)$ is plotted as a function of $k^2$ for
$NN$($^1$S$_0$).
Red and blue lines in Fig.~\ref{fig:sanity}
represent fits to data by the effective range expansion (ERE) at the next-to-leading order (NLO) as
\begin{eqnarray}
k\cot\delta_0(k) &=& \frac{1}{a_0} + \frac{r_0}{2} k^2,
\end{eqnarray}
where $a_0$ and $r_0$ are the scattering length and the effective range, respectively.
  In Fig.~\ref{fig:sanity} (Left),
  inconsistency in ERE parameters is observed:
  The NLO ERE fit obtained from the data at $k^2 <0$ on finite volumes (red line)
  disagrees with the fit to the data at $k^2 >0$ on finite volumes
  together with the infinite volume limit at $k^2<0$ (blue line).
  For the latter fit (the blue line),
  the physical condition of the bound state pole is also violated.
  In Fig.~\ref{fig:sanity} (Right), the NLO ERE fit exhibits a singular behavior
  as the divergent effective range.
See Ref.~\cite{Iritani:2017rlk} for more detailed discussions.

As in the case of  operator dependence, the normality check in Ref.~\cite{Iritani:2017rlk}
indicates that the plateau fitting at $t\simeq 1$ fm suffers from large uncontrolled systematic errors probably due to contaminations from 
the excited states.

\subsection{Source-operator dependence and derivative expansion of the potentials}
\label{subsec:hal}
In Ref.~\cite{Iritani:2018zbt}, we investigated 
 source operator dependence of the  potential as a tool to estimate
the systematics associated with the derivative expansion of the
HAL QCD potential, using two sources, wall and smeared sources.

\begin{figure}[t]
  \centering
  \includegraphics[width=0.47\textwidth,clip]{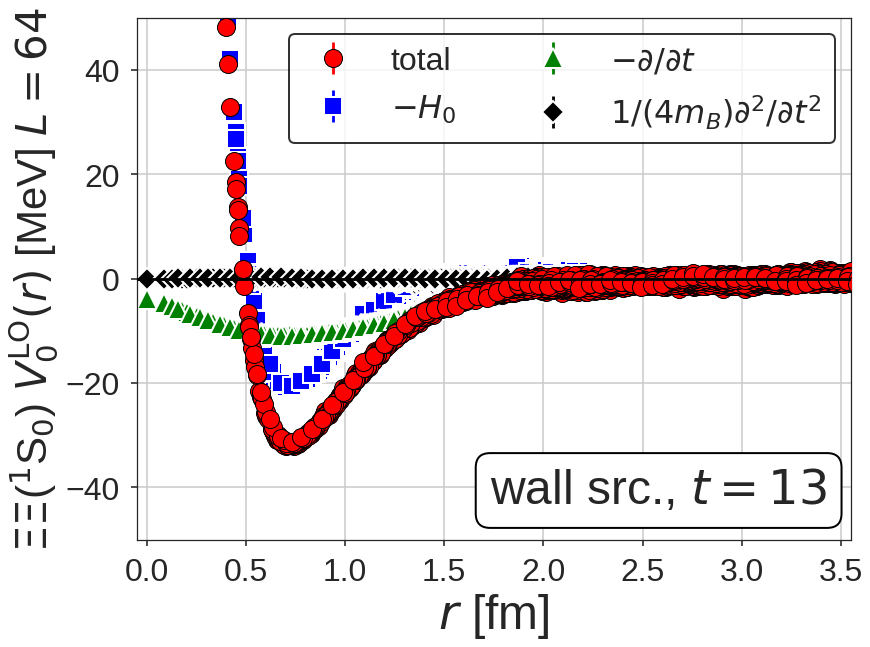}
  \includegraphics[width=0.47\textwidth,clip]{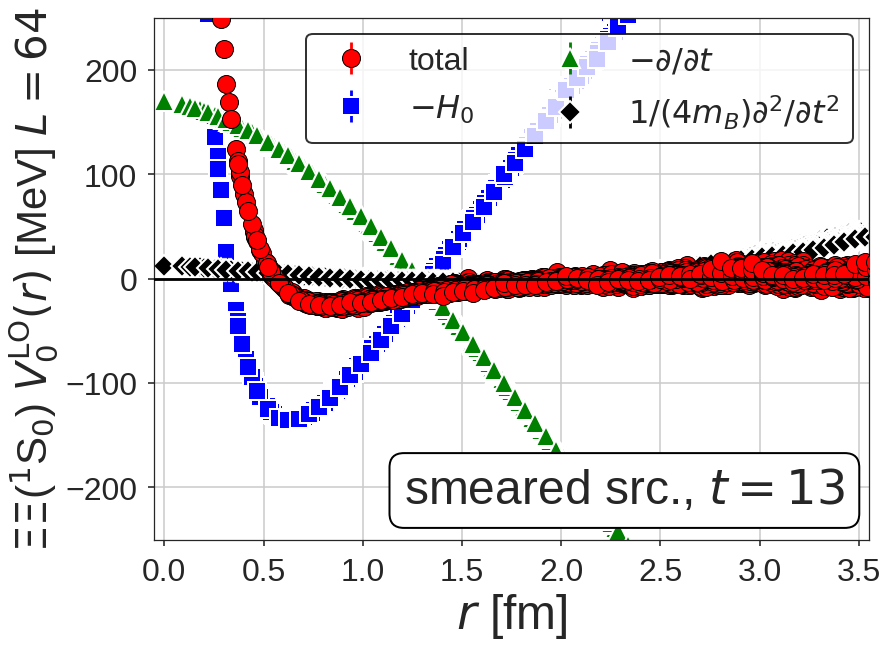}
  \caption{\label{fig:pot_breakup}
    The potential at the leading order (LO) analysis $V_0^{\rm LO}(r)$ (red circles) from the wall source 
    (Left) and the smeared source (Right) for $\Xi\Xi$($^1$S$_0$)  at $t/a=13$ for $L=64$~\cite{Iritani:2018zbt}.    
    The blue squares, green triangles and black diamonds denote 1st, 2nd and 3rd terms in Eq.~(\ref{eq:veff}), respectively.    
  }
\end{figure}

Fig.~\ref{fig:pot_breakup} shows the LO $\Xi\Xi$($^1$S$_0$) potential and its breakup into 1st, 2nd, and 3rd terms in Eq.~(\ref{eq:veff}) 
of the time dependent HAL QCD method from the wall source (Left) and the smeared source (Right).
For the wall source, the 1st term dominates with moderate (negligible) contributions from the 2nd (3rd) term. 
As the 2nd term is not constant as a function of $r$, there exist small but non-negligible contributions from the excited states. 
For the smeared source, on the other hand, all terms are important.
The substantial $r$ dependence of the 2nd term (green triangles), which indicates large contributions from the excited states in the smeared source, is canceled by the 1st term (blue squares) and further corrected by the 3rd term (black diamonds). The total potentials (red circles) from two sources, however,
show qualitatively similar behaviors, which illustrates 
that the time-dependent HAL QCD method works well for extracting the potential irrespective of the source types.  

\begin{figure}[hbt]
  \centering
  \includegraphics[width=0.47\textwidth,clip]{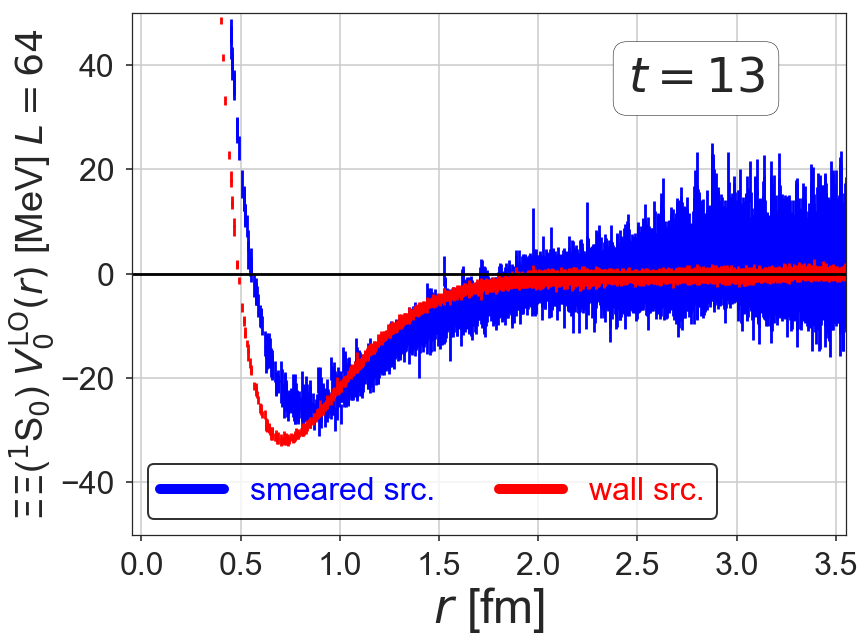}
  \caption{\label{fig:pot_comp}
    A comparison of the LO $\Xi\Xi$($^1$S$_0$) potential $V_0^{\rm LO}(r)$  between the wall source (red) and the smeared source (blue) at $t/a = 13$ \cite{Iritani:2018zbt}.
  }
\end{figure}
\begin{figure}[hbt]
  \centering
  \includegraphics[width=0.49\textwidth,clip]{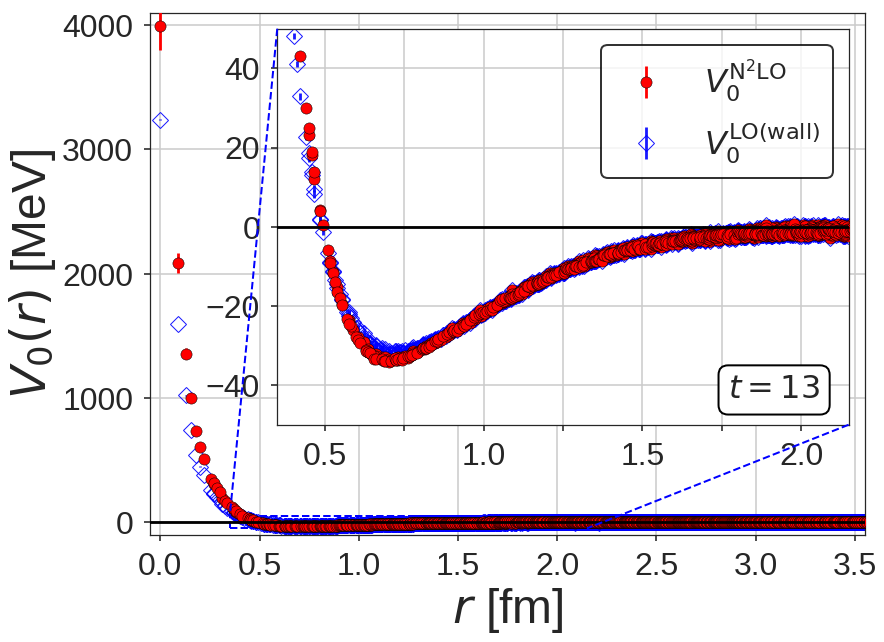}
  \includegraphics[width=0.46\textwidth,clip]{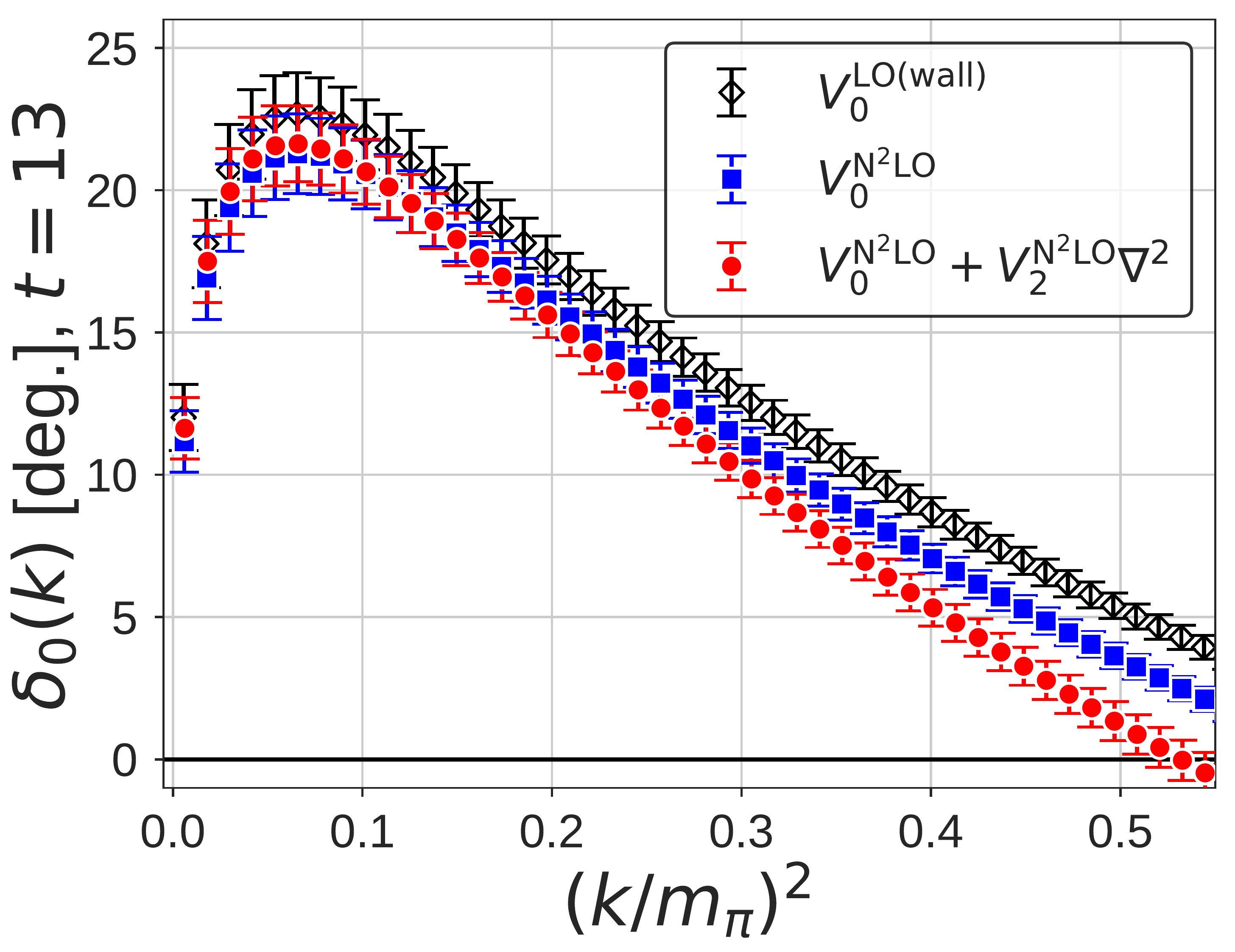}
  \caption{
    \label{fig:vlo_vnlo}
    (Left) 
      The LO $\Xi\Xi$($^1$S$_0$) potential at the N$^2$LO analysis, $V_0^\mathrm{N^2LO}(r)$ (red circles),
      together with the potential at the LO analysis for the wall source, $V_0^\mathrm{LO(wall)}(r)$
      (blue diamonds) at $t/a=13$ on $L=64$.
    (Right) The scattering phase shifts $\delta_0(k)$ from $V_0^{\rm LO(wall)}$ (black diamonds),
    $V_0^{\rm N^2LO}(r)$ (blue squares) and  $V_0^{\rm N^2LO}(r) + V_2^{\rm N^2LO}(r)\nabla^2$ (red circles) at $t/a=13$.
      Both figures from Ref.~\cite{Iritani:2018zbt}.
  }
\end{figure}

Fig.~\ref{fig:pot_comp} shows a comparison of the total potentials between two sources, $V_0^{\rm LO(wall/smear)}(r)$,
at $t/a = 13$.
The potential approaches to zero within errors at larger $r$ for both sources,
indicating that contributions from inelastic states are suppressed.
  While the potentials from two sources are very similar,
  there exists the non-zero difference between them.
  We find that this difference becomes smaller as $t$ increases,
  where the potential from the wall (smeared) source is independent of (dependent on) $t$~\cite{Iritani:2018zbt}.
  This indicates that the effects of higher order terms in the derivative expansion exist
  in the data from the smeared source.
Using the difference, $V_0^{\rm N^2LO}(r)$ and $V_2^{\rm N^2LO}(r)$ in Eq.~(\ref{eq:pot:N2LO})
can be determined. 
In Fig.~\ref{fig:vlo_vnlo} (Left),  $V_0^{\rm N^2LO}(r)$ together with
$V_0^{\rm LO(wall)}(r)$ on $L=64$ at $t/a=13$ are plotted. 
As we find that  $V_0^{\rm N^2LO}(r)$ agrees well with $V_0^{\rm LO(wall)}(r)$ except at short distances,
we expect that $V_0^{\rm LO(wall)}(r)$ works well to reproduce physical observables at low energies.
Indeed, as shown in  Fig.~\ref{fig:vlo_vnlo} (Right), 
the N$^2$LO correction to the S-wave scattering phase shift $\delta_0$ is small at low energies,  showing not only that the derivative expansion converges well but also that the LO analysis for the wall source is sufficiently good at low energies.  

\section{Anatomy: Excited state contaminations in the effective energy shifts}
\label{sec:anatomy}
We now show our main results of this paper, where we analyze the behaviors of $\Xi\Xi$($^1$S$_0$) temporal correlation functions for both wall and smeared sources using the HAL QCD potential
and demonstrate that  contaminations of elastic excited states  cause pseudo-plateau behaviors at early time slices.

The strategy of our analysis is as follows. Provided
that the leading order HAL QCD potential from the wall source is found to be a reasonable approximation of the exact potential,
we evaluate eigenfunctions of the Hamiltonian with this potential in the finite 
box whose eigenvalues are below the inelastic threshold.   
We then calculate overlap factors between these eigenmodes and the $\Xi\Xi$ ($^1$S$_0$) correlation functions, in terms of which we reconstruct pseudo-plateau behaviors of the energy shifts.
We also show that the plateaux of the temporal correlation functions projected to the lowest or the 2nd lowest eigenfunction agree with their eigenvalues. 
This fact demonstrates that both HAL QCD potential method and accurate extraction of energy shifts from the temporal correlation function give consistent results.

\subsection{Eigenvalues and eigenfunctions in the finite box}
\label{subsec:eigens}
We first evaluate eigenvalues and eigenfunctions of the leading-order HAL QCD Hamiltonian in a finite box given by
\begin{eqnarray}
  H^{\rm LO} = H_0 + U, \qquad U\equiv V_0^{\rm LO(wall)} ,
\end{eqnarray}
where we take  $V_0^{\rm LO(wall)}$ on each volume for  $U$, since
$V_0^{\rm LO(wall)}$ is a reasonable approximation of the exact potential $U$.\footnote{In Appendix~\ref{app:n2lo}, we employ the potential at the N$^2$LO analysis instead,
  and find that the results are consistent with the LO results.
  }
Note that the volume dependence of $V_0^{\rm LO(wall)}$ is negligible, thanks to the short-ranged nature of the interaction.

In a finite lattice box, 
eigenvalues and eigenfunctions of the Hermitian matrix $H^{\rm LO}$
can be easily obtained.
The $n$-th eigenvalue of $H^{\rm LO}$ in the $A_1^{+}$ representation below the 
inelastic threshold\footnote{
For the $\Xi\Xi$ system in the $^1S_0$ channel at $m_\pi = 0.51$ GeV, the lowest inelastic threshold is
either $\Xi^\ast\Xi$ or $\Sigma\Omega$ in $^5$D$_0$ channel,
which corresponds to $W_\mathrm{th} - 2m_\Xi \simeq 0.25 - 0.30$~GeV
on $L=64-40$, using
$m_\Sigma = 1.40$~GeV, $m_{\Xi^\ast} = 1.68$~GeV and $m_\Omega = 1.74$~GeV.}
is tabulated in Table~\ref{tab:deltaEn_summary}, where
we show the energy shift compared to the threshold,
$\Delta E_n \equiv W_n - 2m_\Xi$.
The number of excited states below the inelastic threshold is
3, 4 and 6 on $L = 40$, 48, and 64, respectively.
For  larger volume,
the energy gap becomes smaller and the number of elastic excited states increases.
\begin{table}
  \centering
  \begin{tabular}{c|ccc}
\hline
\hline
$\Delta E_n$ [MeV] & $L= 40$ & $L = 48$ & $L = 64$ \\
\hline
$n = 0$  & $-5.5(1.0)\left(^{+1.8}_{-0.4}\right)$  & $-2.8(0.4)\left(^{+1.1}_{-0.1}\right)$  & $-1.5(0.3)\left(^{+0.4}_{-0.1}\right)$ \\
$n = 1$  & $77.2(0.8)\left(^{+0.8}_{-4.7}\right)$  & $52.0(0.4)\left(^{+3.6}_{0.0}\right)$  & $28.4(0.3)\left(^{+0.4}_{-0.1}\right)$ \\
$n = 2$  & $161.5(1.0)\left(^{+0.2}_{-6.8}\right)$  & $110.0(0.5)\left(^{+3.2}_{0.0}\right)$  & $60.4(0.4)\left(^{+0.3}_{-0.1}\right)$ \\
$n = 3$  & $236.5(1.1)\left(^{+1.0}_{-0.7}\right)$  & $164.9(0.6)\left(^{+2.1}_{-0.5}\right)$  & $93.2(0.4)\left(^{+0.7}_{0.0}\right)$ \\
$n = 4$  &  --- & $216.3(0.4)\left(^{+1.0}_{-0.4}\right)$  & $124.1(0.3)\left(^{+0.2}_{-0.1}\right)$ \\
$n = 5$  &  --- & ---  & $155.8(0.3)\left(^{+0.6}_{-0.0}\right)$ \\
$n = 6$  &  --- & ---  & $186.5(0.3)\left(^{+0.8}_{0.0}\right)$ \\
\hline
\hline
\end{tabular}
  \caption{
    Eigenvalues of the $n$-th eigenfunction below the inelastic threshold
    in the $A_1^+$ representation
    of the HAL QCD Hamiltonian $H^{\rm LO}$ in each volume.
    Eigenvalues are given in terms of the energy shifts from the threshold, $\Delta E_n \equiv W_n - 2m_\Xi$.
    Central values and statistical errors are evaluated at $t/a = 13$,
    while the systematic errors are estimated by using the results at $t/a = 14, 15, 16$.
\label{tab:deltaEn_summary}}
\end{table}

\begin{figure}[hbt]
  \centering
  \includegraphics[width=0.47\textwidth,clip]{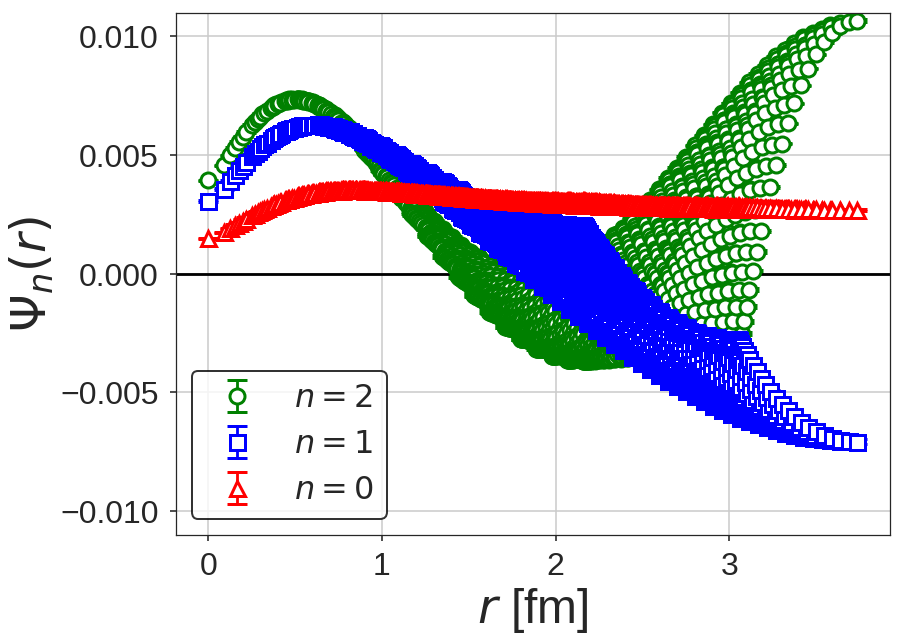}
  \includegraphics[width=0.47\textwidth,clip]{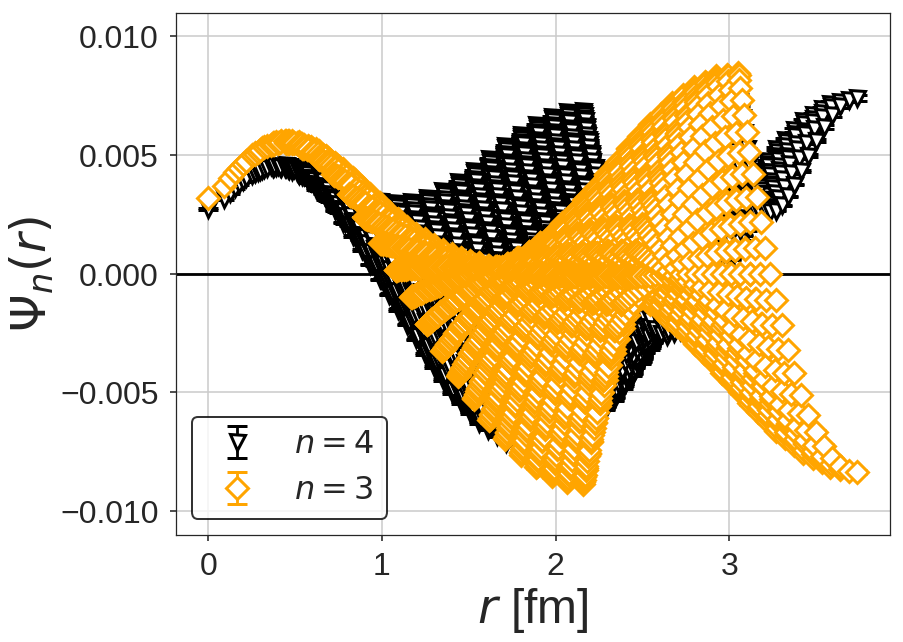}
  \caption{\label{fig:eigen_functions:48}
   Eigenfunctions of elastic states in the $A_1^+$ representation 
      of the HAL QCD Hamiltonian $H^{\rm LO}$  below the inelastic threshold
      $\Psi_n(\vec{r})$
      on $L=48$ at $t/a = 13$
      for $n=0, 1, 2$ (Left) and $n=3, 4$ (Right),
    where red up-pointing triangles, blue squares, 
    green circles, orange diamonds and black down-pointing triangles correspond to 
    $n=0, 1, 2, 3$ and 4, respectively.  The eigenfunction is normalized as $\sum_{\vec{r}} |\Psi_n(\vec{r})|^2 = 1$ and $\Psi_n(\vec{0}) > 0$.
    Errors are statistical only.
  }
\end{figure}

Fig.~\ref{fig:eigen_functions:48} shows
the eigenfunctions $\Psi_n(\vec{r})$ on $L=48$,
with
$\sum_{\vec{r}} |\Psi_n(\vec{r})|^2 = 1$ and $\Psi_n(\vec{0}) > 0$.%
\footnote{
  $\Psi_n(\vec{r})$ is a multi-valued function of $r$ due to the
    effect of the (periodic) finite box.}
Up to a normalization,
$\Psi_n(\vec{r})$ corresponds to the NBS wave function $\psi^{W=W_n}(\vec{r})$ in a finite volume.
The lowest state $\Psi_0(\vec{r})$ has the similar shape to the $R$-correlator for the wall source,
which shows a weak peak structure around $r \lesssim 1$~fm and becomes flat at large distances without any nodes,
while the eigenfunctions for the excited states have nodes,
whose number increases as the eigenvalue becomes larger.
The short distance structures for $\Psi_{n>0}(\vec{r})$, which has a steeper peak around $r < 1$~fm than that of $\Psi_0(\vec{r})$ resemble the $R$-correlator for the smeared source.
Eigenfunctions on other volumes are collected
 in Appendix~\ref{app:eigen_func}.

\subsection{Decomposition of the $R$-correlator via eigenfunctions}
\label{subsec:decomposition}

\begin{figure}[hbt]
  \centering
  \includegraphics[width=0.47\textwidth,clip]{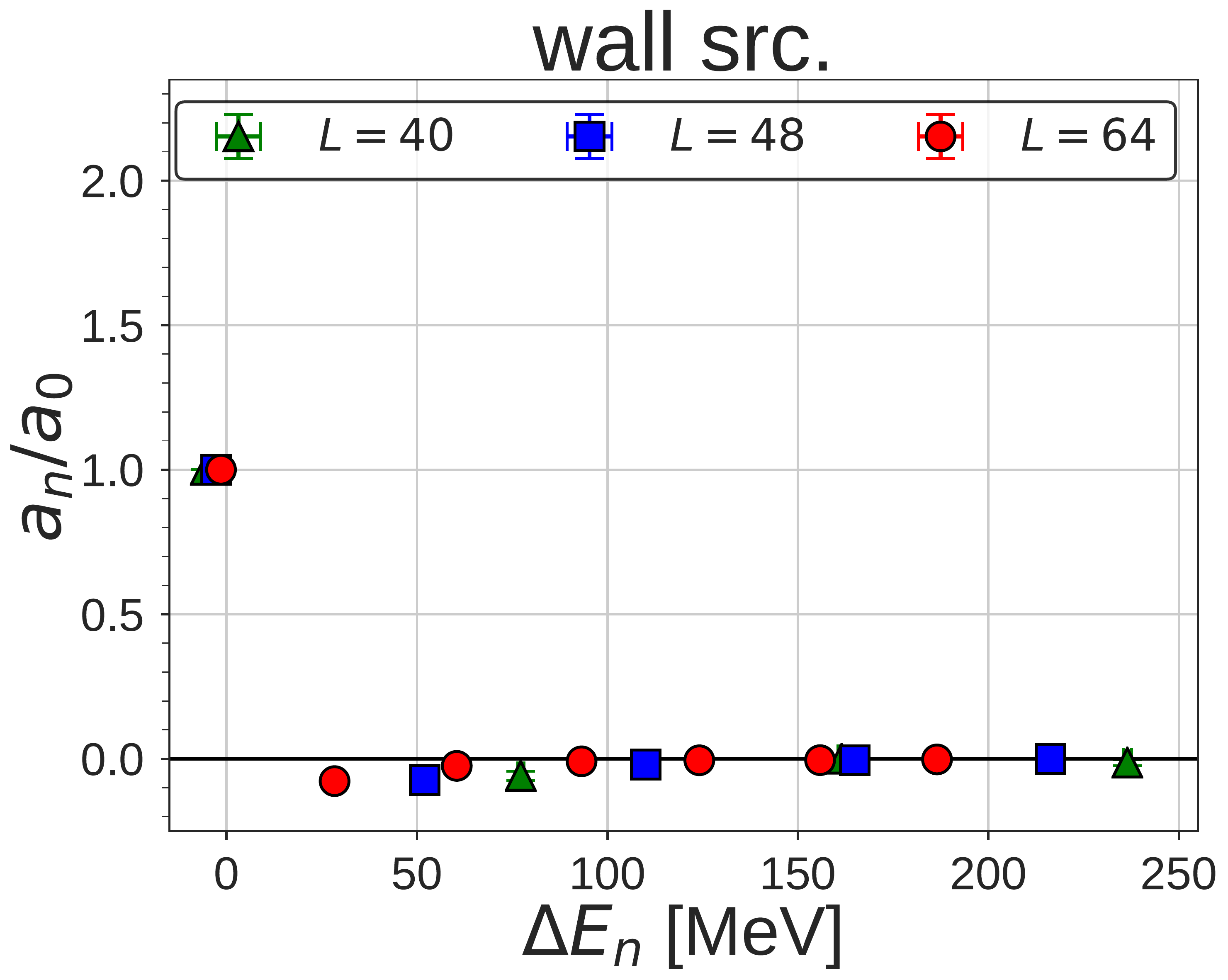}
  \includegraphics[width=0.47\textwidth,clip]{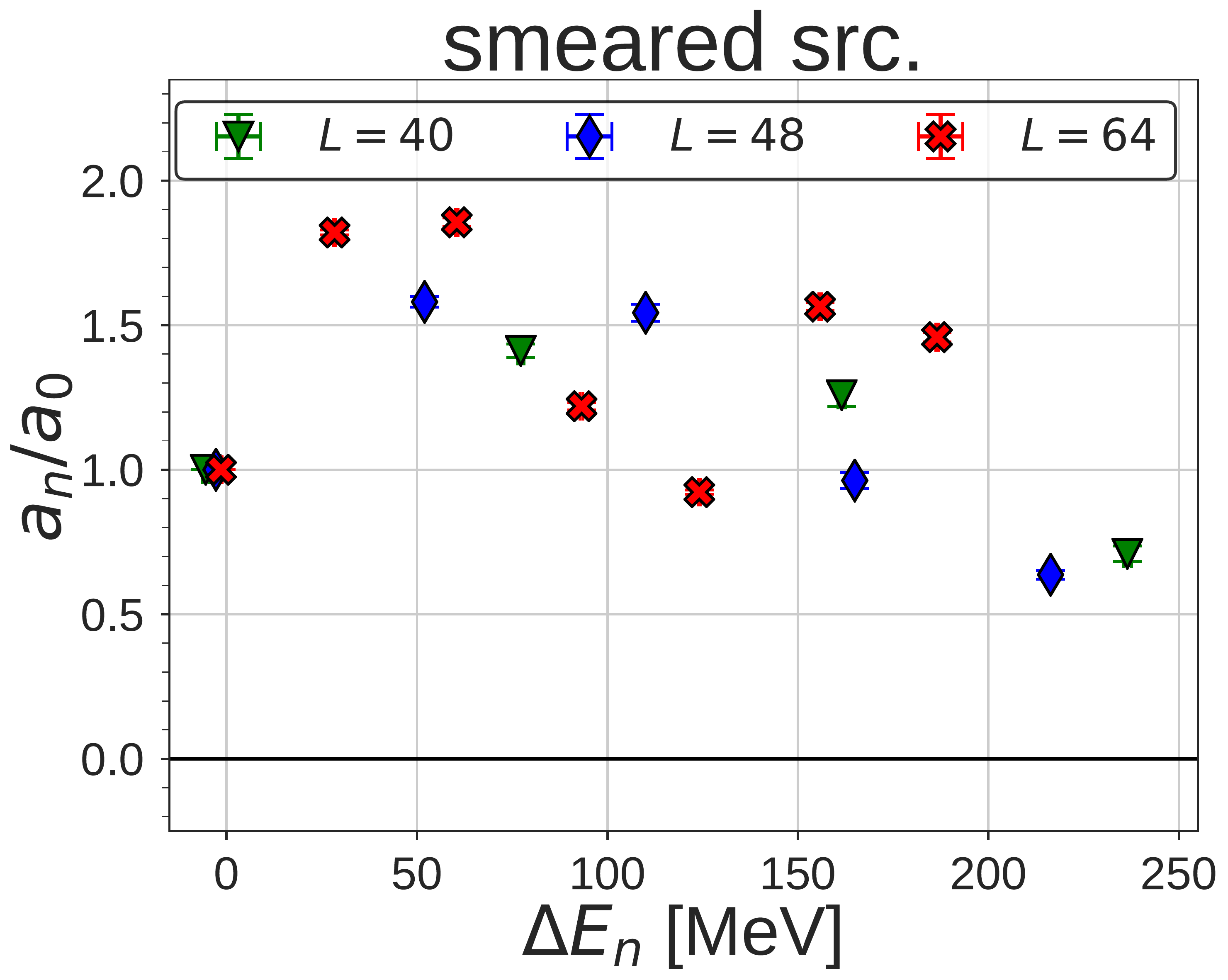}
  \caption{
    \label{fig:an_a0}
    The ratio of the overlap coefficients $a_n/a_0$ in the $R$-correlator
     obtained at $t/a = 13$ for the wall source (Left) and the smeared source (Right)
    on three volumes.
  }
\end{figure}

Since the $R$-correlator is dominated by elastic states at moderately large $t$,
we can expand it in terms of eigenfunctions of $H^\mathrm{LO}$ as
\begin{equation}
  R^\mathrm{wall/smear}(\vec{r}, t) = \sum_n a_n^\mathrm{wall/smear} \Psi_n(\vec{r})
  e^{-\Delta E_n t},
  \label{eq:Rcorr_decomp}
\end{equation}
where the overlap coefficient $a_n$ characterizes the magnitude of the contribution from the corresponding eigenfunction.
Using the orthogonality of $\Psi_n(\vec{r})$, $a_n$ can be determined by
\begin{eqnarray}
  a_n^\mathrm{wall/smear} = \sum_{\vec{r}} \Psi_n^\dag(\vec{r}) R^\mathrm{wall/smear}(\vec{r}, t) e^{\Delta E_n t}.
  \label{eq:Rcorr_decomp:a_n}
\end{eqnarray}
The magnitude of the corresponding excited state contamination in the $R$-correlator
is represented by the ratio $a_n/a_0$.
In Fig.~\ref{fig:an_a0},
we plot $a_n/a_0$ obtained at $t/a = 13$
as a function of $\Delta E_n$
for the wall source (Left) and the smeared source (Right).
Calculations  at $t/a=14, 15, 16$ 
confirm that the results are almost independent of $t$ within statistical errors,
indicating that the decomposition is reliable.\footnote{
  In Appendix~\ref{app:inelastic}, we check how well 
  the decomposition in Eq.~(\ref{eq:Rcorr_decomp}) approximates the original $R$-correlator. 
  It is found that the magnitude of the residual relative to the original $R$-correlator
  at $t/a = 13$  is  as small as
  ${\cal O}(10^{-5}-10^{-6})$ for the wall source
  and
  $0.4$\%, $2$\%, $5$\% for the smeared source  on $L=40$, $48$, $64$, respectively. 
}
In the case of the wall source, $R^\mathrm{wall}(r,t)$ has a large overlap with the ground state,
and $|a_{n>0}/a_0|$ is smaller than 0.1.
In the case of the smeared source, on the other hand,
$|a_n/a_0| \sim \mathcal{O}(1)$ and thus all  elastic excited states significantly contribute to the $R$-correlator.

\begin{figure}[hbt]
  \centering
  \includegraphics[width=0.47\textwidth,clip]{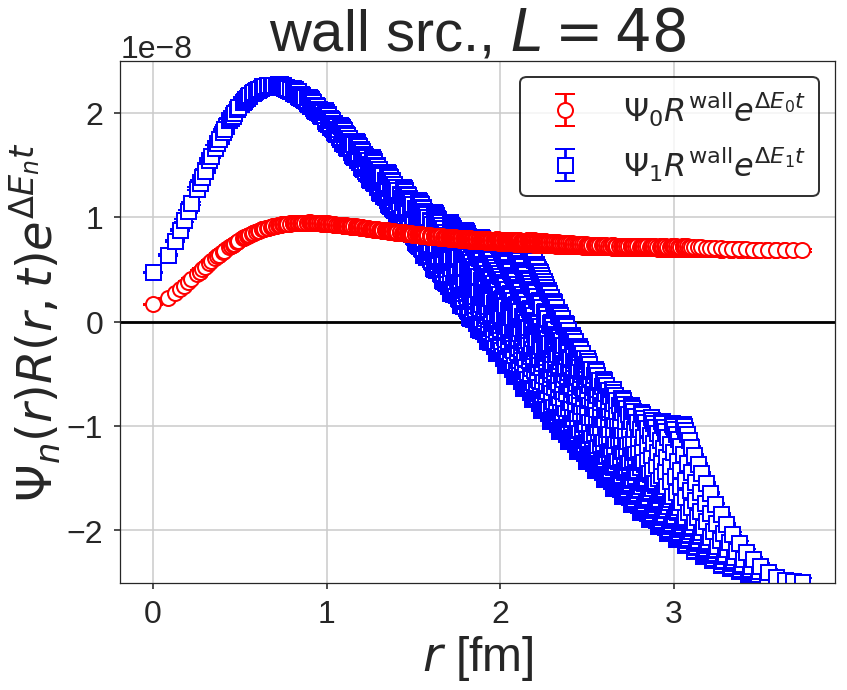}
  \includegraphics[width=0.47\textwidth,clip]{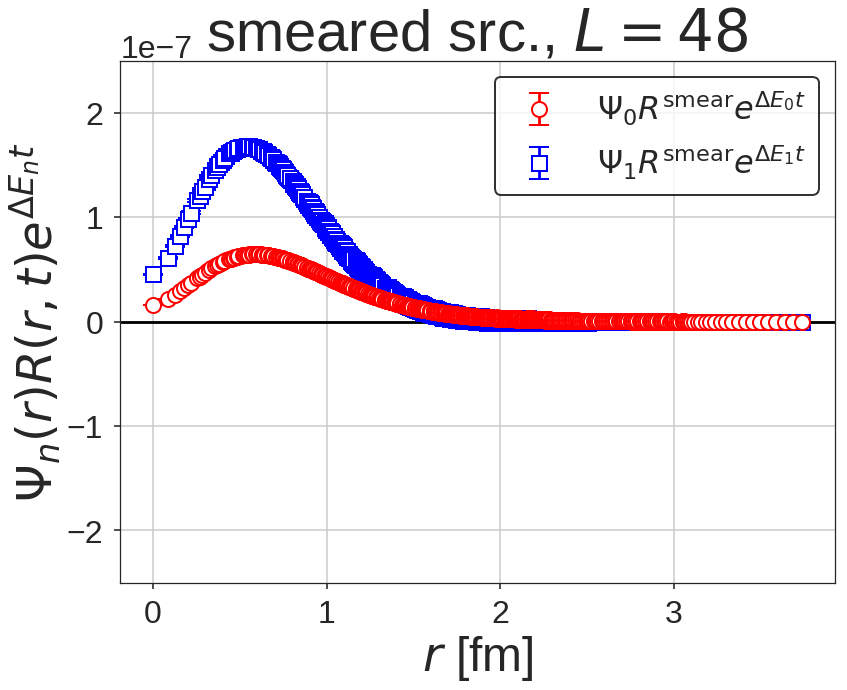}
  \caption{
    \label{fig:Roverlap}
    The overlap between the $R$-correlator and the eigenfunction,  
    $\Psi_n^\dag(\vec{r}) R(\vec{r},t) e^{\Delta E_n t}$, as a function of $r$ at $t/a=13$ on $L=48$
    for the ground state ($n=0$, red circles) and the first excited state ($n=1$, blue squares)
    in the case of the wall source (Left) and the smeared source (Right).
  }
\end{figure}

This difference of the magnitude in $a_n/a_0$ between two sources 
can be understood 
from the overlap between the $R$-correlator and the eigenfunctions.
Fig.~\ref{fig:Roverlap} shows
the spatial profile of the overlap, $\Psi_n^\dag(\vec{r}) R(\vec{r},t) e^{\Delta E_n t}$,
for the wall source (Left) and the smeared source (Right) on $L = 48$,
whose spatial summation corresponds to $a_n$.
The contribution from the first excited state (blue squares) is highly suppressed
for the wall source, thanks to the cancellation
between positive values at short distances and
negative values at long distances
of $\Psi_{n=1}^\dag(\vec{r})R^\mathrm{wall}(\vec{r},t) e^{\Delta E_{n=1} t}$.
For the smeared source, on the contrary, the contamination from the 1st excited state remains non-negligible, 
due to the absence of the negative parts in $\Psi_{n=1}^\dag(\vec{r})R^\mathrm{smear}(\vec{r},t) e^{\Delta E_{n=1} t}$.

In the literature, the smeared source is often employed in the direct method, in order to
suppress contributions from (inelastic) excited states in a ``single-baryon'' correlation function.
The same smeared source, however, does not necessarily reduce the contaminations from the
elastic excited  states in the two-baryon correlation function,
as is explicitly shown in Fig.~\ref{fig:an_a0}.
Indeed, one of the most relevant parameters which control the magnitudes of elastic state contributions
is the relative distance $\vec r$ between two baryons at the source, which appears as
\begin{eqnarray}
  \frac{1}{L^3}\sum_{\vec x} B(\vec x) B(\vec x + \vec r) 
  =  \sum_{\vec p} \tilde B(\vec p) \tilde B(-\vec p) e^{i \vec p \cdot \vec r}, \qquad
  \tilde B(\vec p) \equiv \frac{1}{L^3} \sum_{\vec x} B(\vec x) e^{-i\vec p\cdot \vec x}
\end{eqnarray}
in the center of mass system.
Almost all  literature of the direct method have employed the smeared source 
essentially corresponding to $|\vec{r}|=0$, 
which
implies that elastic states for all $\vec p$ are equally generated at the source. 
Thus the choice $|\vec{r}|=0$ (or $\vert\vec{r}\vert\ll 1$)  is one of the possible reasons for large contaminations  
from elastic excited states in the case of the smeared source.
As long as $\vert\vec r\vert$ is non-zero and large, however, 
modes with non-zero $\vec{p}$ may be suppressed due to the oscillating factor $e^{i \vec p \cdot \vec r}$. 
\footnote{
  Ref.~\cite{Berkowitz:2015eaa} reported  
  the discrepancy in the effective energies 
  between  the zero  displaced ($|\vec r|=0$)  and the non-zero  displaced ($|\vec r|\not=0$) source operators,
  which can be naturally understood from this viewpoint,
  rather than the existence of two bound states claimed in Ref.~\cite{Berkowitz:2015eaa}.}
  The study in which the momentum projection is performed for each baryon in the source operator
  is recently reported in Ref.~\cite{Francis:2018qch}.

The temporal correlation function $R(t)$ is reconstructed in terms of eigenfunctions as
\begin{eqnarray}
  R^\mathrm{wall/smear}(t) &=& 
 \sum_{\vec{r}} R^\mathrm{wall/smear}(\vec{r}, t) 
  = \sum_{\vec{r}} \sum_n a_n^\mathrm{wall/smear}\Psi_n(\vec{r})
  e^{-\Delta E_n t} 
  = \sum_n b_n^\mathrm{wall/smear} e^{-\Delta E_n t} , ~~~~~~~
  \label{eq:bn_factor}
\end{eqnarray}
where
$b_n \equiv a_n \sum_{\vec{r}}\Psi_n(\vec{r})$,
whose ratio $b_n/b_0$ gives the
magnitude of the contamination to $R(t)$ from the $n$-th elastic excited  state.

\begin{figure}[hbt]
  \centering
  \includegraphics[width=0.47\textwidth,clip]{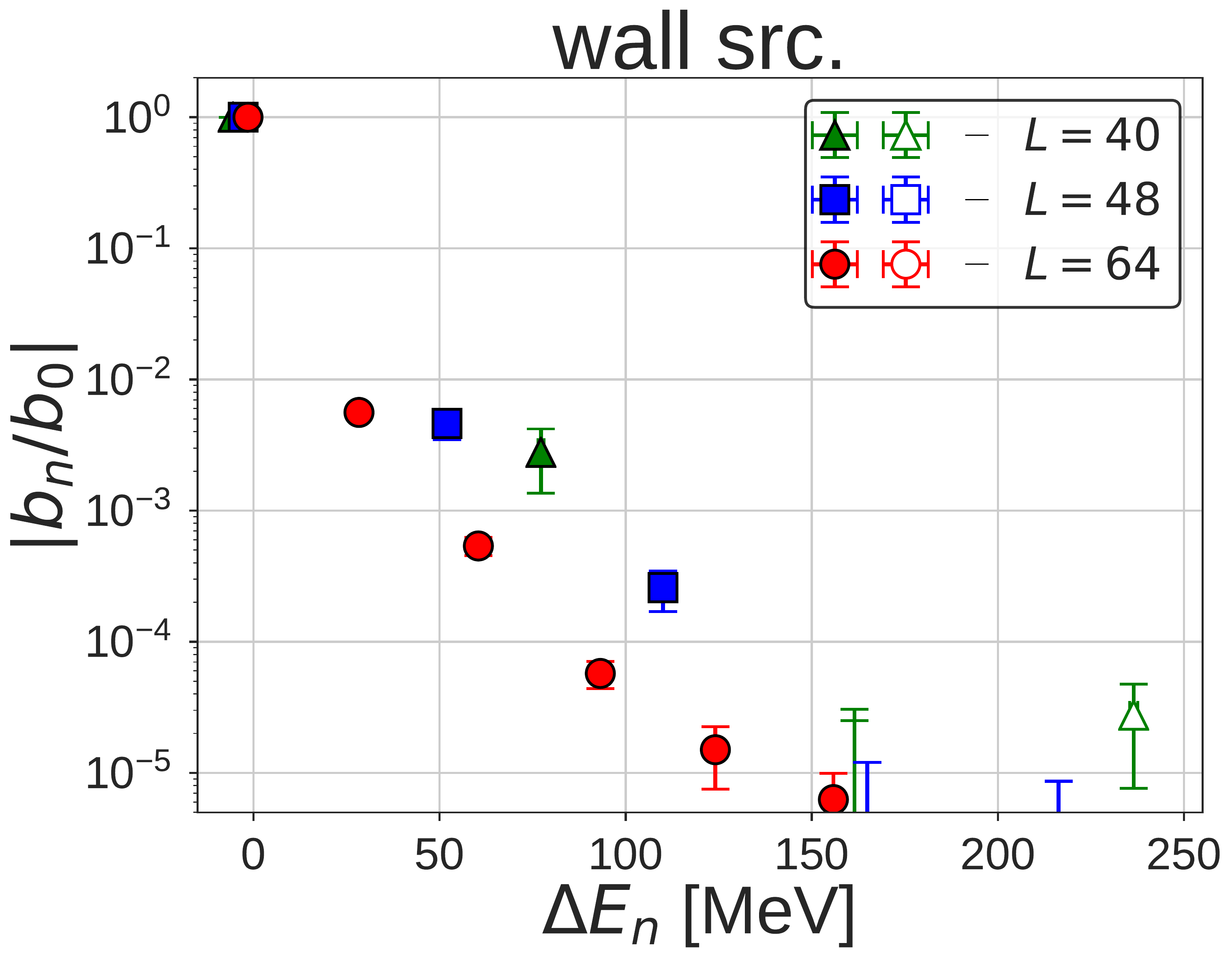}
  \includegraphics[width=0.47\textwidth,clip]{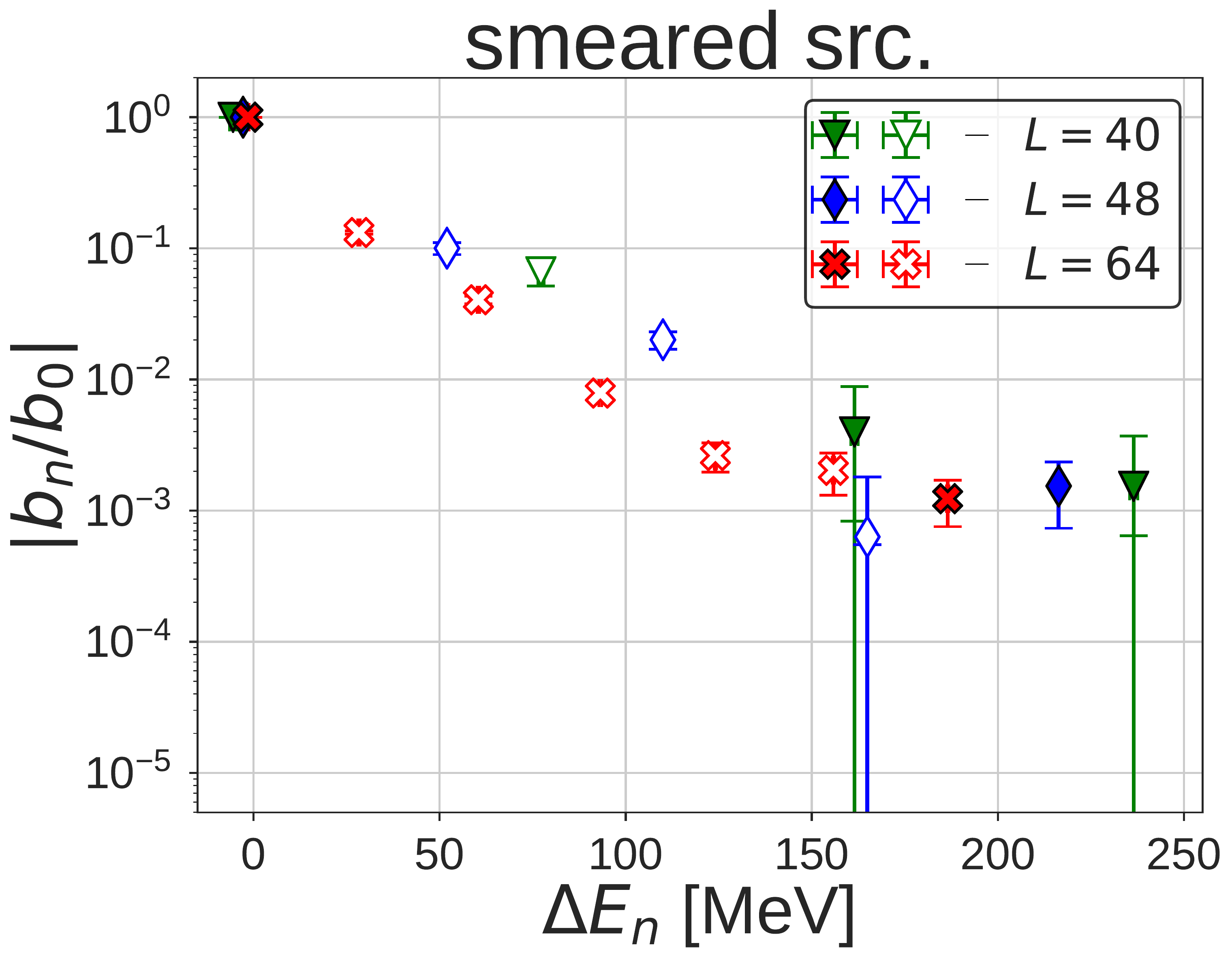}
  \caption{
    \label{fig:bn_b0}
    The ratio of the overlap coefficients in the temporal correlation function
    $|b_n/b_0|$ obtained at $t/a = 13$ for the wall source (Left) and the smeared source (Right)
    on various volumes.
Solid (open) symbols correspond to positive (negative) values of $b_n/b_0$.
  }
\end{figure}

Fig.~\ref{fig:bn_b0} shows $\vert b_n/b_0\vert $ obtained at $t/a=13$  as a function of $\Delta E_n$ on three volumes
for the wall source (Left) and the smeared source (Right).
Solid (open) symbols correspond to positive (negative) values for $b_n/b_0$.
For the wall source, the contamination from the first excited state is found to be smaller than 1\%,
and 
$|b_n/b_0|$ is further suppressed exponentially for higher excited states. 
In the case of the smeared source, 
the contamination from the first excited state is as large as $\sim 10$\% with a negative sign
and the contamination remains to be $\sim 1$\% even for the higher excited state with $\Delta E_n \sim 100$~MeV.

\subsection{Reconstruction of the effective energy shift}
\label{subsec:reconstruction}
\begin{figure}[hbt]
  \centering
  \includegraphics[width=0.47\textwidth,clip]{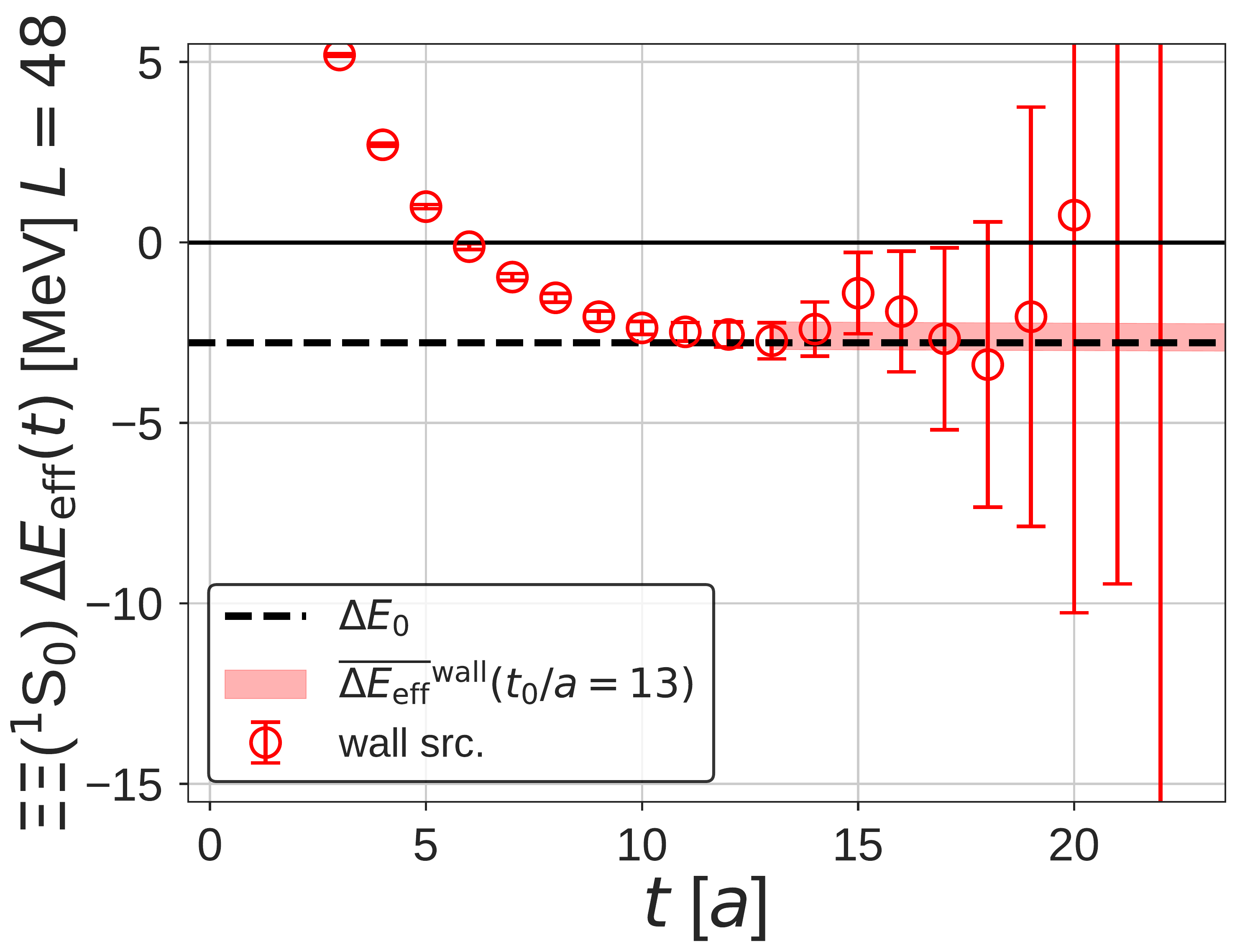}
  \includegraphics[width=0.47\textwidth,clip]{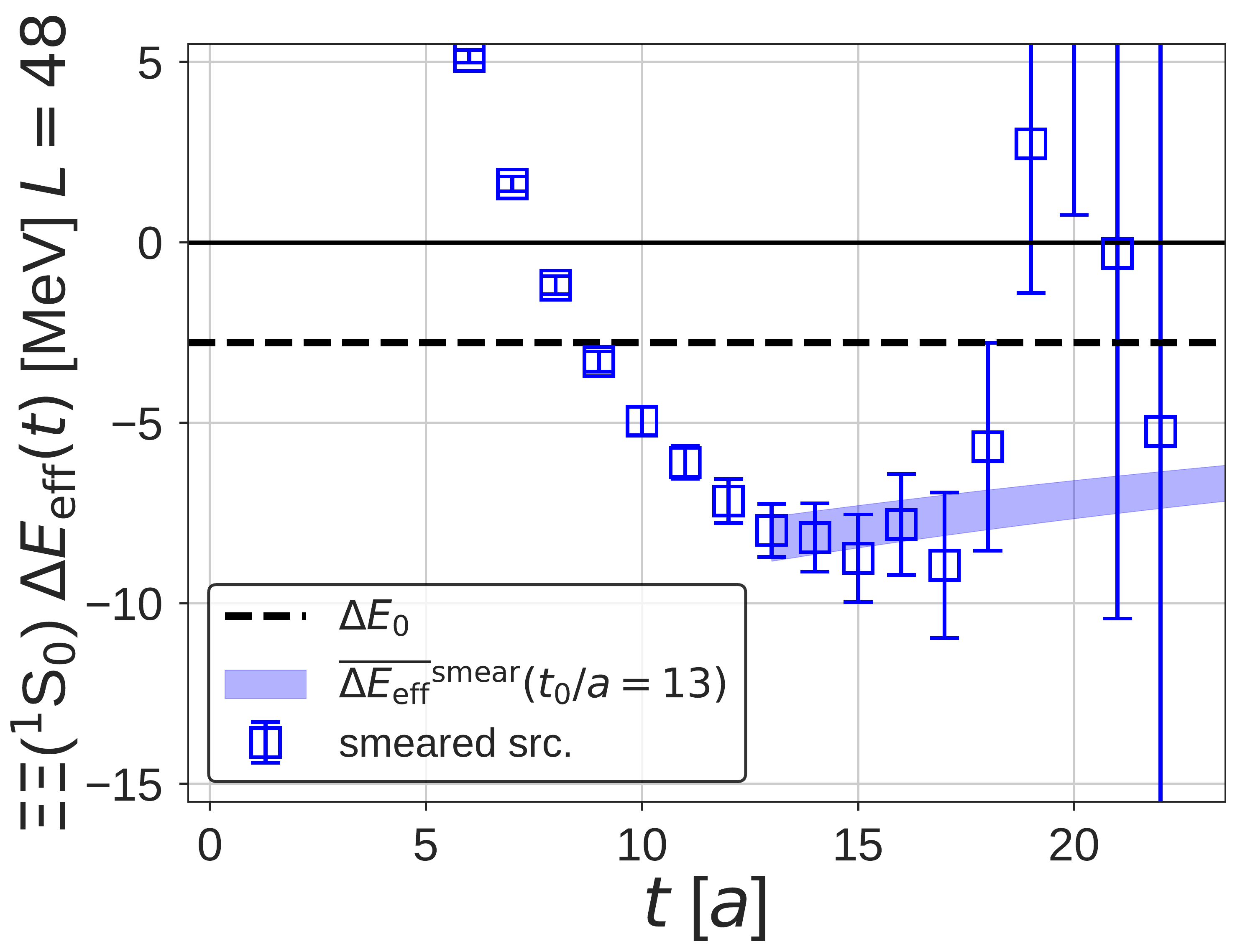}
  \caption{
    \label{fig:ReDEeffComp:48}
    The reconstructed effective energy shifts $\overline{\Delta E}_\mathrm{eff}(t,t_0=13a)$
    with statistical errors are plotted as a function of $t$ (colored bands),
    while the direct measurement of the effective energy shifts from $R$-correlators are plotted by red circles or blue squares.
    The black dashed lines correspond to the energy shift $\Delta E_0 (t_0=13a)$ for the ground state of the
    HAL QCD Hamiltonian $H^{\rm LO}$ in the finite volume. 
 The results on $L=48$ for the wall source (Left) and the smeared source (Right).
  }

\end{figure}

\begin{figure}[hbt]
  \centering
  \includegraphics[width=0.7\textwidth,clip]{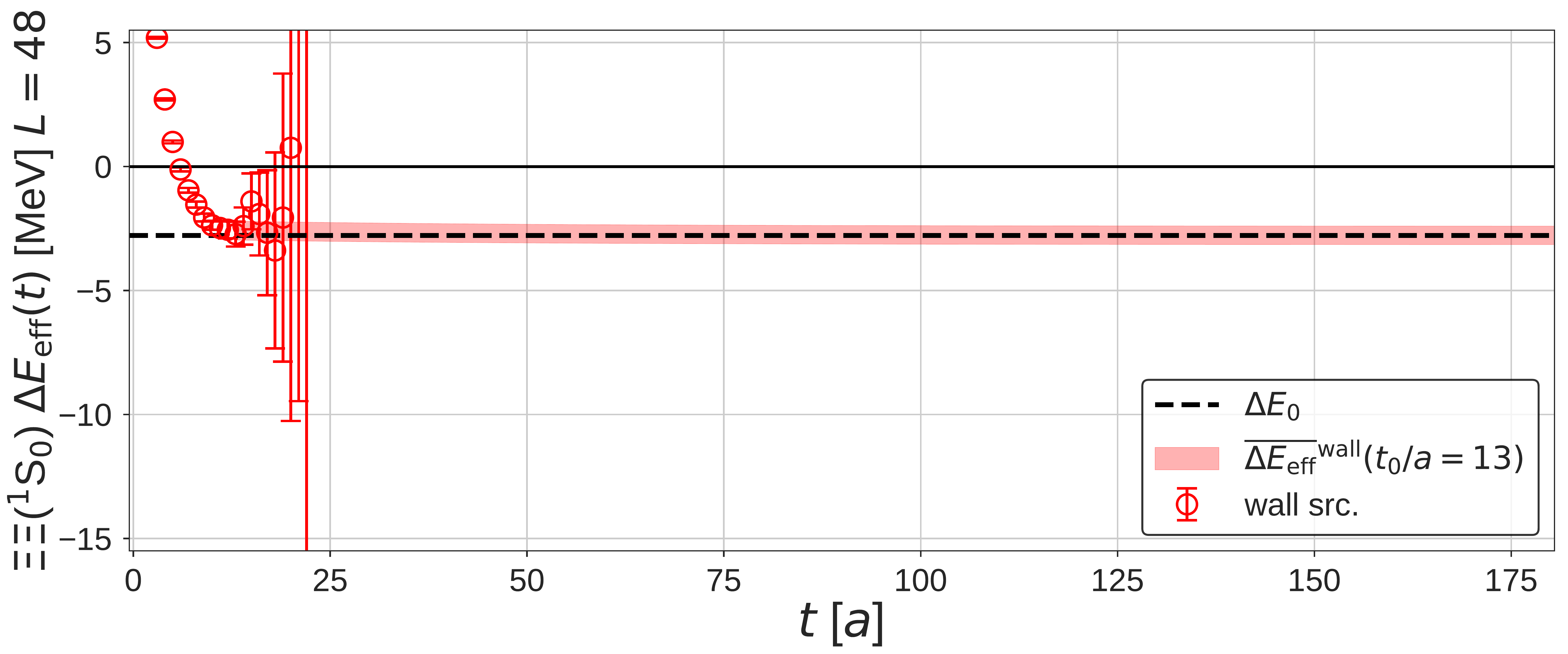}

  \includegraphics[width=0.7\textwidth,clip]{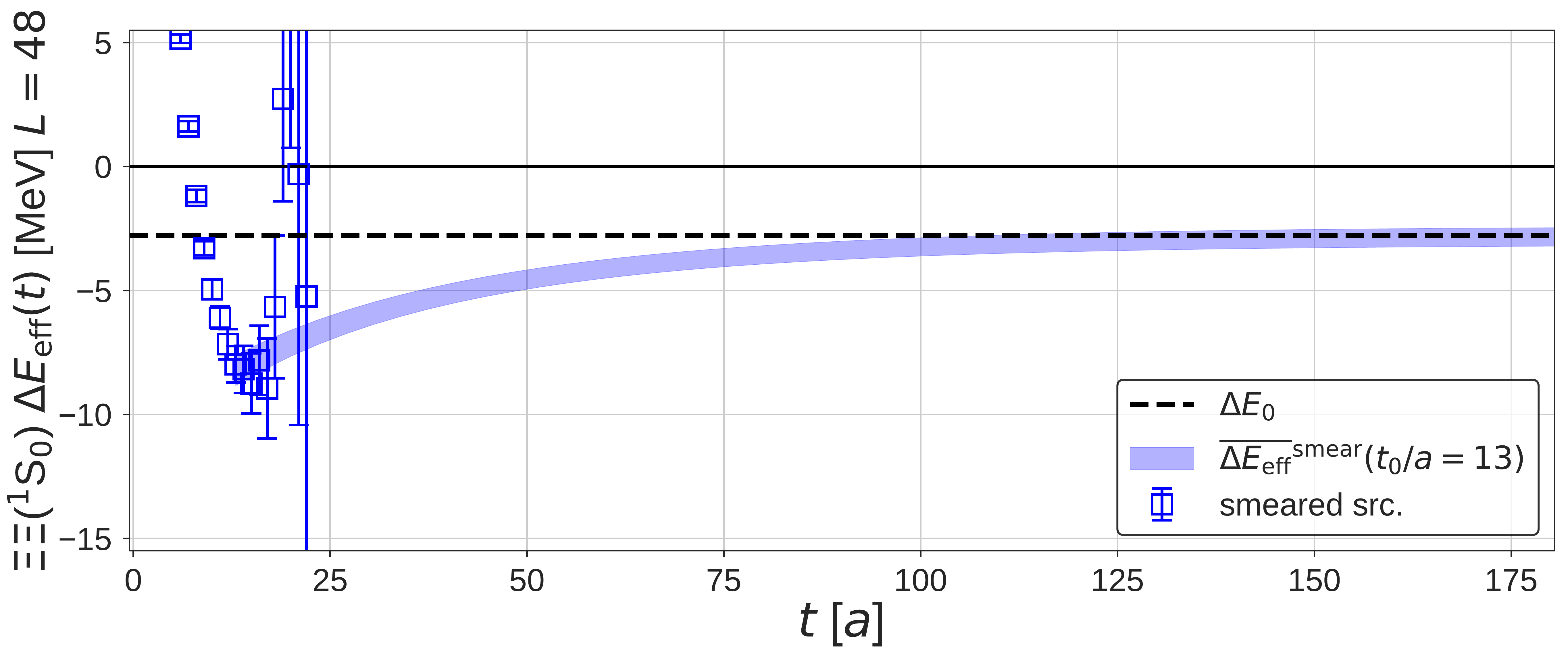}
  \caption{
    \label{fig:ReDEeffComp:48:large_t}
    The same as Fig.~\ref{fig:ReDEeffComp:48}
    but for the wider range of the Euclidean time $t$.
  }
\end{figure}

Let us now examine  the energy shifts obtained from the reconstructed $R$-correlators;
\begin{eqnarray}
  \overline{\Delta E}_\mathrm{eff}^{\mathrm{wall/smear}}(t, t_0)
  = \frac{1}{a}
  \log \left[
  \frac{\displaystyle\sum_{n=0}^{n_\mathrm{max}} b_n^\mathrm{wall/smear} (t_0)e^{-\Delta E_n(t_0)t}}
       {\displaystyle\sum_{n=0}^{n_\mathrm{max}} b_n^\mathrm{wall/smear}(t_0) e^{-\Delta E_n(t_0)(t+a)}}
       \right] ,
  \label{eq:ReDEeff}
\end{eqnarray}
where we take $n_\mathrm{max} = 3$, 4, 6 for $L = 40$, 48, 64, respectively,
corresponding to the number of elastic excited  states below the inelastic threshold, and
$b_n(t_0)$ and $\Delta E_n(t_0)$ are extracted at fixed $t_0$.

In Fig.~\ref{fig:ReDEeffComp:48},
we show the reconstructed effective energy shifts $\overline{\Delta E}_\mathrm{eff}(t, t_0=13a)$,
together with numerical data of the effective energy shifts $\Delta E_\mathrm{eff}(t)$ from the $R$-correlators,
for the wall source (Left) and the smeared source (Right) on $L=48$.
The bands correspond to $\overline{\Delta E}_\mathrm{eff}(t, t_0=13a)$
with statistical errors coming from those of $b_n$ and $\Delta E_n$ at $t_0/a = 13$,
while red circles or blue squares correspond to $\Delta E_\mathrm{eff}(t)$ obtained directly from the $R$-correlator in Sec.~\ref{subsec:direct}.
Here we do not consider $\overline{\Delta E}_\mathrm{eff}(t,t_0=13a)$ for $t/a < 13$,
where inelastic contributions are expected to be larger.
Shown together by the black dashed line represents
the energy shift $\Delta E_0 (t_0=13a)$ for the ground state of the HAL QCD Hamiltonian $H^{\rm LO}$ on $L=48$.

We find that the results of the direct method,
most notably the plateau-like structures around $t/a = 15$,
are well reproduced by $\overline{\Delta E}_\mathrm{eff}(t,t_0)$ for both wall and smeared sources,
indicating that the behaviors of $\Delta E_\mathrm{eff}(t)$ at this time interval in the direct method are explained by the contributions from the several low-lying states.
These plateau-like structures, however, do not necessarily correspond to the plateau of the ground state.
Indeed, in the case of the smeared source, there is a clear discrepancy between
the value of the plateau-like structure and the eigenvalue $\Delta E_0$ of the ground state.
This is a consequence of large excited state contaminations in
the correlation function for the smeared source.
In the case of the wall source, on the other hand, since the overlap with the ground state is large, 
the value of the plateau-like structure is consistent with the value $\Delta E_0$.

The fate of the plateau-like structures is
more clearly seen in Fig.~\ref{fig:ReDEeffComp:48:large_t},
where we plot the behaviors at asymptotically large $t$ of
the reconstructed effective energy shifts $\overline{\Delta E}_\mathrm{eff}(t, t_0=13a)$
for the wall source (red band) and the smeared source (blue band).
While the plateau-like structure at $t/a \sim 15$ for the wall source is almost unchanged at larger $t$,
the value of $\overline{\Delta E}_\mathrm{eff}(t, t_0=13a)$ in the case of the smeared source
gradually changes as $t$ increases
until it reaches to the value of the ground state, $\Delta E_0$\blue{,} at $t/a \sim 100$.

In Appendix~\ref{subapp:delta_E_r:vol}, we perform the same analysis on other volumes
and observe essentially the same behaviors as in the case of $L=48$:
For the wall source,
the value of the plateau-like structure at $t/a \sim 15$
remains almost unchanged at larger $t$
and is consistent with $\Delta E_0$.
For the smeared source,
the value of the plateau-like structure at $t/a \sim 15$
is inconsistent with $\Delta E_0$.
The deviation is found to be larger on a larger volume,
due to severer contaminations from the excited states  
on larger volumes (See Sec.~\ref{subsec:decomposition}).\footnote{
  Values of pseudo-plateaux do not strongly depend on volumes, while the correct values $\Delta E_0$ do. 
  This is a counter example against the argument in Refs.~\cite{Wagman:2017tmp,Beane:2017edf} in which it is claimed that
  the volume-independence of the plateaux  guarantees their correctness.} 
The value of $\overline{\Delta E}_\mathrm{eff}(t, t_0=13a)$ for the smeared source
gradually changes at $t$ increases
and the ground state saturation is realized at
$t/a \gtrsim 50, 100$ and $150$
or
$t \gtrsim 5, 10$ and $15$ fm 
on $L = 40, 48$ and $64$, respectively.
These time scales for the ground state saturation are actually not surprising but rather natural,
considering the fact that the lowest excitation energy is as small as
$\delta E \equiv \Delta E_1 - \Delta E_0 \simeq 84$, 55 and 30~MeV on $L = 40$, 48 and 64, respectively.

These results clearly reveal that
the plateau-like structures at $t/a \sim 15$ for the smeared source
are pseudo-plateaux caused by contaminations of the elastic excited states.\footnote{
  In Appendix~\ref{subapp:cut_off},
  we show that the dominant contamination comes from the first excited state.}
While the effective energy shifts from the wall source happen to be saturated by the ground state even at $t/a \sim 15$,
it is generally difficult to confirm  that a plateau-like structure corresponds to the correct energy  shift of the ground state  without the help of other inputs,
such as the HAL QCD potential analysis in the present case.
Since the calculation of the energy shift from the $R$-correlator at $t \sim (\delta E)^{-1}$ 
is impractical due to the exponentially growing noises,
one cannot obtain the correct spectra
from the plateau identification in the direct method unless sophisticated variational techniques~\cite{Luscher:1990ck} are employed.
\footnote{
Application of the variational method to two-baryon systems has started lately
but mostly with respect to the flavor space~\cite{Francis:2018qch}.
 It will be interesting to perform the variational method
with respect to relative coordinate space for two baryons in addition.
In the lattice study of meson-meson scatterings~\cite{Briceno:2017max},
the danger of the excited state contaminations in the plateau fitting
has been already recognized and the use of the variational method is known to be mandatory.
}

\subsection{Projection with improved sink operator}
\label{subsec:eigen-proj}

Once the finite-volume eigenmodes of  $H^{\rm LO}$
with the HAL QCD potential are known,
an improved two-baryon sink operator for a designated eigenstate can be constructed as
\begin{eqnarray}
  \mathcal{J}_{BB}^{\rm sink}(t) &=& \sum_{\vec{r}} \Psi_n^\dag(\vec{r}) \sum_{\vec{x}} B(\vec{x}+\vec{r},t)B(\vec{x},t) ,
    \label{eq:EeffProj}
\end{eqnarray}
which is expected to have a large overlap to the $n$-th elastic state.\footnote{Use of such an improved operator as a ``source'' requires 
additional calculations and  thus is left for future study.}
This is equivalent to considering the generalized temporal correlation function
with the choice of $g(\vec{r}) = \Psi_n^\dag(\vec{r})$ in Eq.~(\ref{eq:gen_op}),
\begin{eqnarray}
  R^{(n)} (t) &\equiv& \sum_{\vec{r}} \Psi_n^\dag(\vec{r}) R(\vec{r}, t) ,
\end{eqnarray}
from which we define the effective energy shift for the $n$-th eigenfunction as
\begin{equation}
  \Delta E_\mathrm{eff}^{(n)}(t) = \frac{1}{a} \log \frac{R^{(n)}(t)}{R^{(n)}(t+a)} .
\end{equation}

\begin{figure}[hbt]
  \centering
  \includegraphics[width=0.47\textwidth,clip]{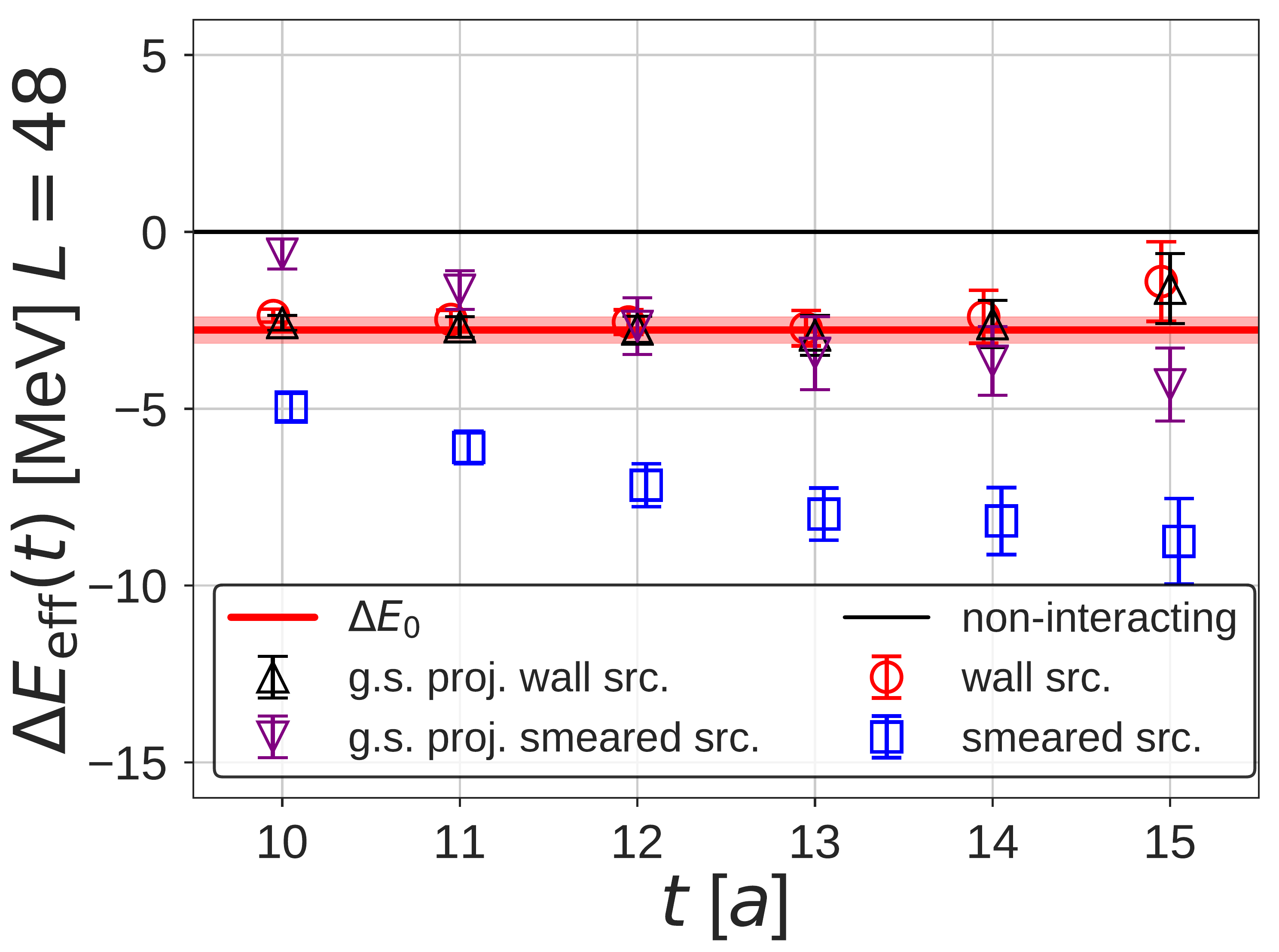}
  \includegraphics[width=0.47\textwidth,clip]{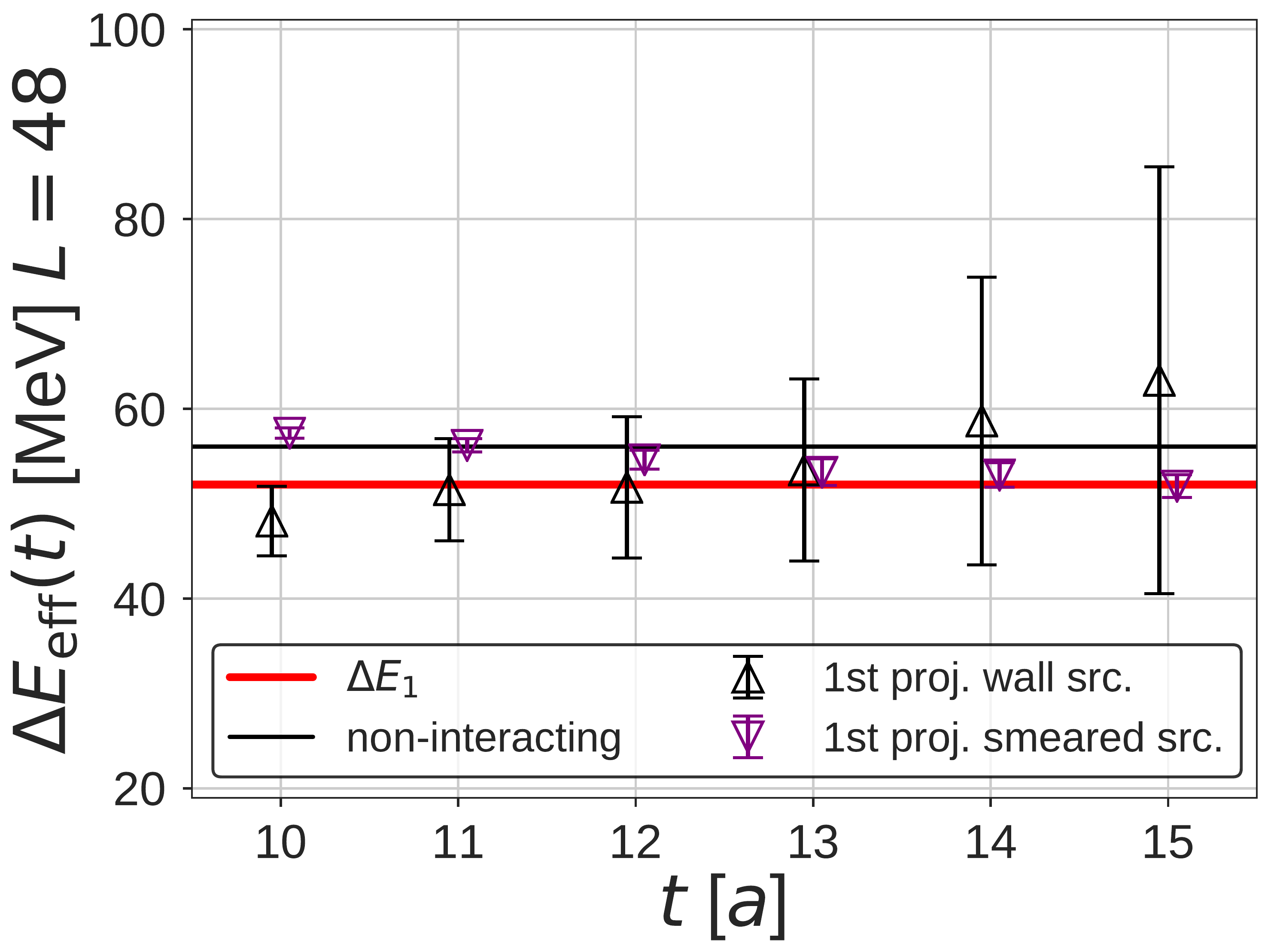}
  \caption{
    \label{fig:Eeff_proj:48}
    The effective energy shift $\Delta E_\mathrm{eff}^{(n)}(t)$ on $L=48$
    from the wall source (black up triangles) and the smeared source (purple down triangles)
    for the ground state (Left) and the first excited state (Right).
    Red bands represent the energy shifts with statistical errors obtained from the HAL QCD Hamiltonian $H^{\rm LO}$, 
    while black lines represent those for a non-interacting system.
    In the left figure, the effective energy shifts in the direct method
    for the wall source (red circles) and the smeared source (blue squares) are also shown.
  }
\end{figure}

Fig.~\ref{fig:Eeff_proj:48} shows the effective energy shift $\Delta E_\mathrm{eff}^{(n)}(t)$
using the wall source (black up-triangles) and the smeared sources (purple down-triangles)
on $L=48$ for the ground state (Left) and the first excited state (Right).
Shown together are 
the energy shift ($\Delta E_0$ or $\Delta E_1$) 
with statistical errors (red bands),
obtained from $H^{\rm LO}$  with the HAL QCD potential $V_0^\mathrm{LO(wall)}(r)$ at $t/a=13$,
as well as that for a non-interacting system (black lines).
In the case of the ground state (Fig.~\ref{fig:Eeff_proj:48} (Left)),
the effective energy shifts from the direct method for the wall source (red circles)
and the smeared source (blue squares) are also plotted for comparisons.

First of all, after the sink projections, 
the results with the wall source and those with the smeared source agree well
 around $t/a \sim 13$
not only for the ground state but also for the first excited state.
This is in sharp contrast with the fact that the results in the direct method without
projections disagree between two sources for the ground state.
Although a small overlap with the  first excited state causes
relatively large statistical errors in the case of the wall source, 
an agreement between two sources for the first excited states is rather striking,
and serves as a non-trivial check for the reliability of the effective energy shifts with the sink projection.
Moreover, results after the sink projections also agree with those from the HAL QCD Hamiltonian.
Although the sink projection utilizes the information of the HAL QCD potential through eigenfunctions, 
agreements in effective energy shifts within statistical errors for both ground and  first excited states 
provide a non-trivial consistency check between the HAL QCD method and the direct method with proper projection.
 In other words, results in Fig.~\ref{fig:Eeff_proj:48} establish that
  (i)  the HAL QCD potential correctly describes the energy shifts of two baryons in the finite volume for both ground and excited states, and that
  (ii)  these energy shifts can also be extracted in the direct method if and only if  interpolating operators are highly improved.
Since the origins of systematic uncertainties are
generally quite different between the two methods, 
such a ``projection check''
would be useful in future lattice QCD studies for two-baryon systems.

In recent years, it has been argued that seemingly  inconsistent results for the $NN$ systems
 at  heavy pion masses  between L\"uscher's finite volume method and the HAL QCD method
may indicate some theoretical deficits in one of the two methods.
It is now clear from our analysis that 
L\"uscher's method and the HAL QCD method agree quantitatively with each other, as it should be so theoretically.

\section{Summary}
\label{sec:summary}

  In our previous works~\cite{Iritani:2016jie, Iritani:2017rlk},
  it has been shown that the plateau fitting
  of the eigenenergies at early Euclidean times $t$,
  employed in  the direct method,
  is generally unreliable for multi-baryon systems,
  due to the appearance of pseudo-plateaux caused
   by contaminations of the excited states with small gap corresponding to the 
   elastic scattering states  on the finite volume. 
     In this paper, 
   we  quantified the degree of contaminations from such excited states
   by decomposing
  the two-baryon correlation functions in terms of 
  the finite-volume eigenmodes of   the HAL QCD Hamiltonian.
 
By taking  $\Xi\Xi$ ($^1$S$_0$) system at $m_\pi = 0.51$~GeV
  in (2+1)-flavor lattice QCD with the wall and smeared quark sources for
   $La = 3.6, 4.3, 5.8$~fm,
    we showed  that the excited state contaminations 
  are suppressed for the wall source,
  while those for the smeared source are substantial and become severer
  on a larger spatial extent. 
  For the smeared source, the plateau-like structures at $t = 1 \sim 2$~fm
  are shown to be  pseudo-plateaux and the plateau with the ground state
   saturation is realized only at  $t > 5 \sim 15$~fm corresponding to the inverse of the lowest excitation energy.
  We also demonstrated that one can optimize
  the two-baryon operator
  utilizing the finite-volume eigenmode of the HAL QCD Hamiltonian.
  The effective energies from the temporal correlation functions with the optimized operators
  are found to be consistent with the finite volume spectra obtained from the HAL QCD Hamiltonian.
  This result establishes not only that
   the correct finite-volume spectra can be accessed by  employing
   highly optimized operators even in the direct method
   but also that the HAL QCD method and the direct method
   agree in the finite volume spectra for the two baryon systems.
  Thus the long-standing issue on the consistency
  between L\"uscher's finite volume method and the HAL QCD method is positively resolved
  at least for the particular system considered here.
  The next step is to carry out comprehensive studies of baryon-baryon
 interactions around the physical quark masses  in the HAL QCD method, which are partly underway (see, e.g.~\cite{Gongyo:2017fjb,Iritani:2018sra,Inoue:2018axd}).
  Those will reveal not only the nature of exotic dibaryons but also the equation of state of dense baryonic matter.

\acknowledgments
We thank the authors of Ref.~\cite{Yamazaki:2012hi} and ILDG/JLDG~\cite{conf:ildg/jldg, Amagasa:2015zwb}
for providing the gauge configurations.
Lattice QCD codes of
CPS~\cite{CPS}, Bridge++~\cite{bridge++} and the modified version thereof by Dr. H.~Matsufuru, 
cuLGT~\cite{Schrock:2012fj} and domain-decomposed quark solver~\cite{Boku:2012zi,Teraki:2013}
are used in this study.
The numerical calculations have been performed on BlueGene/Q and SR16000 at KEK, HA-PACS at University of Tsukuba,
FX10 at the University of Tokyo and K computer at RIKEN R-CCS (hp150085, hp160093).
This work is supported in part by the Japanese Grant-in-Aid for Scientific
Research (No. JP24740146, JP25287046, JP15K17667, JP16K05340, JP16H03978, JP18H05236, JP18H05407),
by MEXT Strategic Program for Innovative Research (SPIRE) Field 5,
by a priority issue (Elucidation of the fundamental laws and evolution of the universe) to be tackled by using Post K Computer,
and by Joint Institute for Computational Fundamental Science (JICFuS).

\clearpage
\appendix

\section{The finite volume spectra
  from the N$^2$LO  potential}
\label{app:n2lo}

In the main text (Sec.~\ref{sec:anatomy}), we study the finite volume spectra 
 to be used for the spectral decomposition of the $R$-correlator 
using the LO potential. 
In this appendix,
 we employ the potential at the N$^2$LO analysis (Eq.~(\ref{eq:pot:N2LO}))
  and examine a stability of the finite volume spectra against the order of the derivative expansion.
The HAL QCD Hamiltonian with the N$^2$LO potential is given by
\begin{equation}
  H^\mathrm{N^2LO}  = H_0 + V_0^\mathrm{N^2LO}(r) + V_2^\mathrm{N^2LO}(r)\nabla^2.
  \label{eq:H_N2LO}
\end{equation}
Note that, while $H$ is non-Hermitian,
its eigenenergies are real
since the eigen equation can be rewritten as the definite generalized Hermitian eigenvalue problem~\cite{Iritani:2018zbt}.
The eigenenergies at $L = 64$ using $V_0^\mathrm{LO(wall)}$, $V_0^\mathrm{N^2LO}$,
and $V_0^\mathrm{N^2LO}+V_2^\mathrm{N^2LO}\nabla^2$ are summarized in Table~\ref{tab:N2LO_eigen}.
The results from the different potentials are consistent with each other at low energies
within statistical errors
and the N$^2$LO correction remains to be small (at most $\sim 1$\% difference)
even for higher energies.
 These results are in line with the observation
that the LO analysis from the wall source gives the correct phase shifts at low energies
and the N$^2$LO analysis gives only small correction even at higher energies~\cite{Iritani:2018zbt}.

\begin{table}[h]
  \centering
  \begin{tabular}{c|ccc}
    \hline 
    \hline
$\Delta E_n$ [MeV] & $V_0^\mathrm{LO(wall)}$ & $V_0^\mathrm{N^2LO}$ & $V_0^\mathrm{N^2LO} + V_2^\mathrm{N^2LO} \nabla^2$ \\
 \hline
$n=0$ & -1.5(0.3)   & -1.4(0.3)   & -1.4(0.3)  \\
$n=1$ & 28.4(0.3)   & 28.6(0.3)   & 28.7(0.3)  \\
$n=2$ & 60.4(0.4)   & 60.8(0.4)   & 61.2(0.4)  \\
$n=3$ & 93.2(0.4)   & 93.4(0.4)   & 93.7(0.4)  \\
$n=4$ & 124.1(0.3)   & 124.3(0.3)   & 124.6(0.3)  \\
$n=5$ & 155.8(0.3)   & 156.5(0.3)   & 157.7(0.4)  \\
$n=6$ & 186.5(0.3)   & 187.3(0.3)   & 189.1(0.4)  \\
    \hline
    \hline
  \end{tabular}
  \caption{The finite volume spectra using 
    $V_0^\mathrm{LO(wall)}$, $V_0^\mathrm{N^2LO}$, and $V_0^\mathrm{N^2LO}+V_2^\mathrm{N^2LO}\nabla^2$ at $L = 64$
    evaluated at $t/a=13$.}
  \label{tab:N2LO_eigen}
\end{table}

\clearpage
\section{Eigenfunctions on various volumes}
\label{app:eigen_func}

In this appendix, we show the shapes of eigenfunctions on various volume.

\begin{figure}[h]
  \centering
  \includegraphics[width=0.47\textwidth,clip]{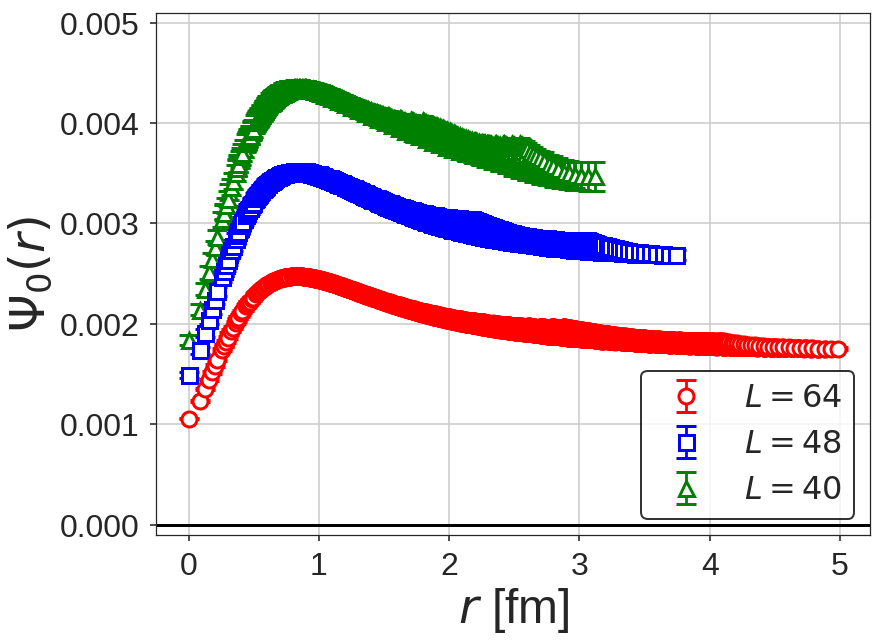}
  \includegraphics[width=0.47\textwidth,clip]{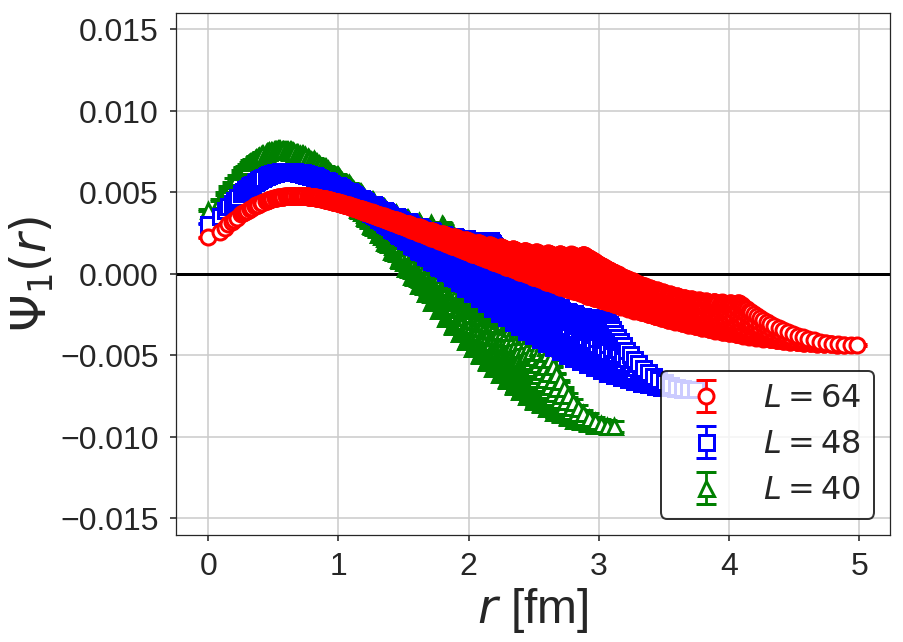}
  \includegraphics[width=0.47\textwidth,clip]{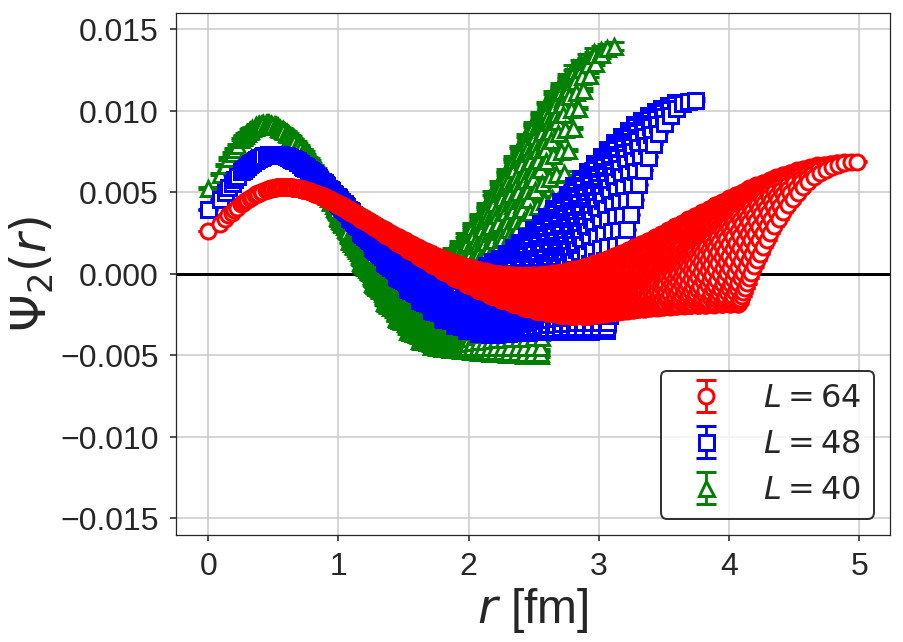}
  \includegraphics[width=0.47\textwidth,clip]{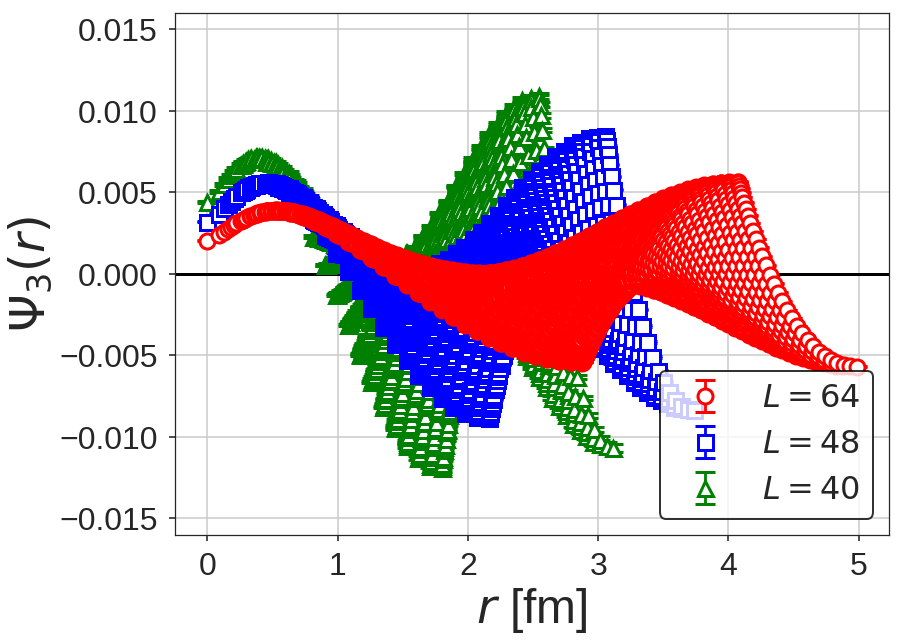}
  \caption{\label{fig:eigen_functions} The low-lying eigenfunctions $\Psi_n(\vec{r})$
    in the $A_1^+$ representation
    of the HAL QCD Hamiltonian with $V_0^\mathrm{LO(wall)}$ at $t/a = 13$
    for $L = 40, 48$ and 64.
    The top-left, top-right, bottom-left, bottom-right figure corresponds to
      $n=0, 1, 2, 3$, respectively.
    The eigenfunction is normalized as $\sum_{\vec{r}} |\Psi_n(\vec{r})|^2 = 1$ and 
   its sign is fixed to satisfy
    $\Psi_n(\vec{0}) > 0$.
  }
\end{figure}

\clearpage
\section{Reconstruction of the $R$-correlator}
\label{app:inelastic}

\begin{figure}[h]
  \centering
  \includegraphics[width=0.48\textwidth,clip]{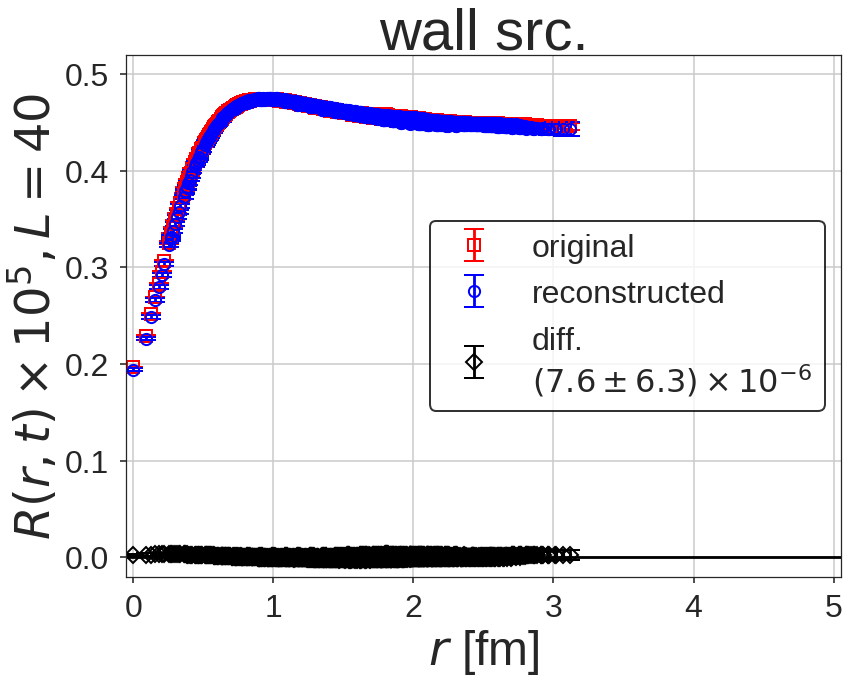}
  \includegraphics[width=0.48\textwidth,clip]{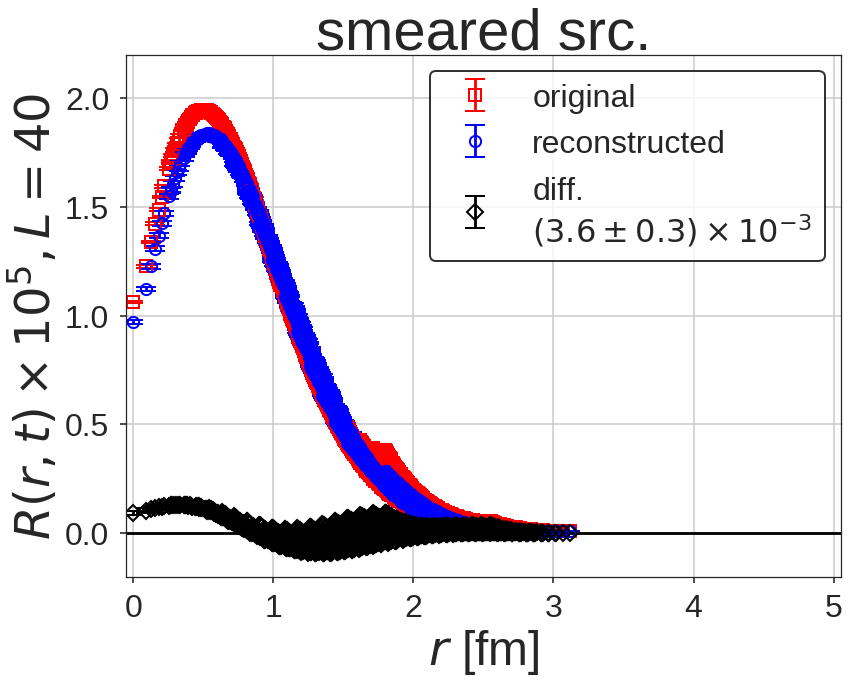}
  \includegraphics[width=0.48\textwidth,clip]{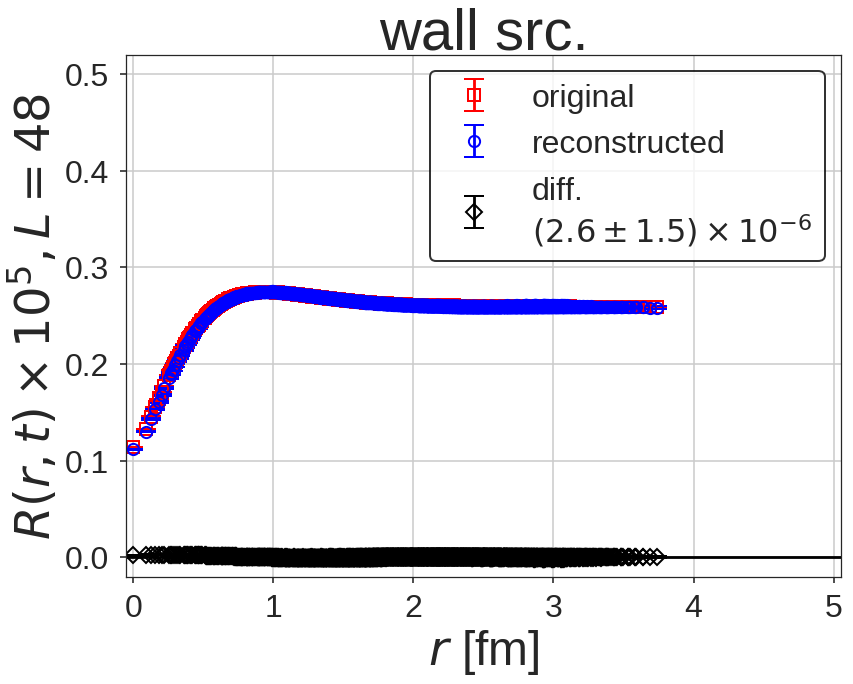}
  \includegraphics[width=0.48\textwidth,clip]{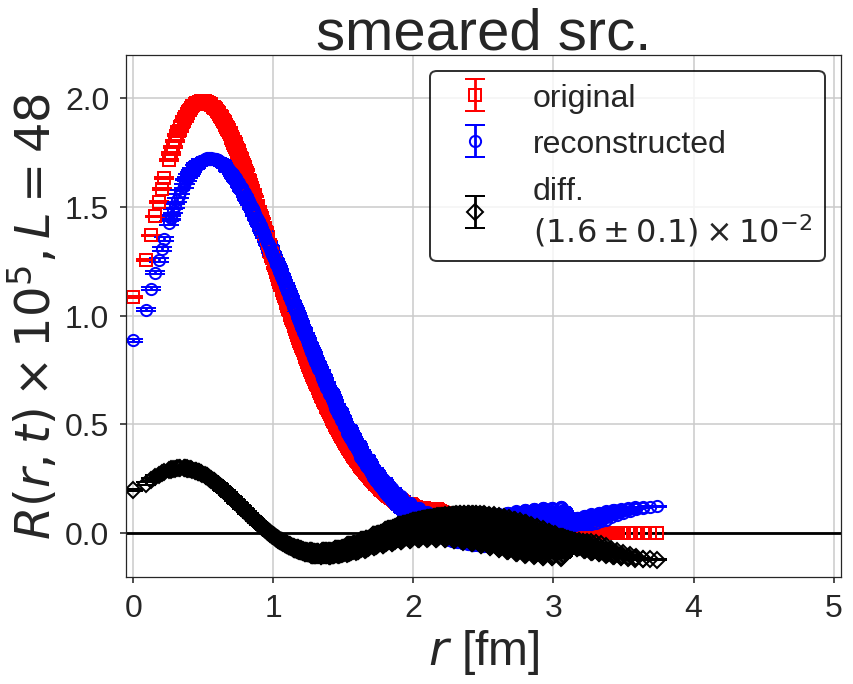}
  \includegraphics[width=0.48\textwidth,clip]{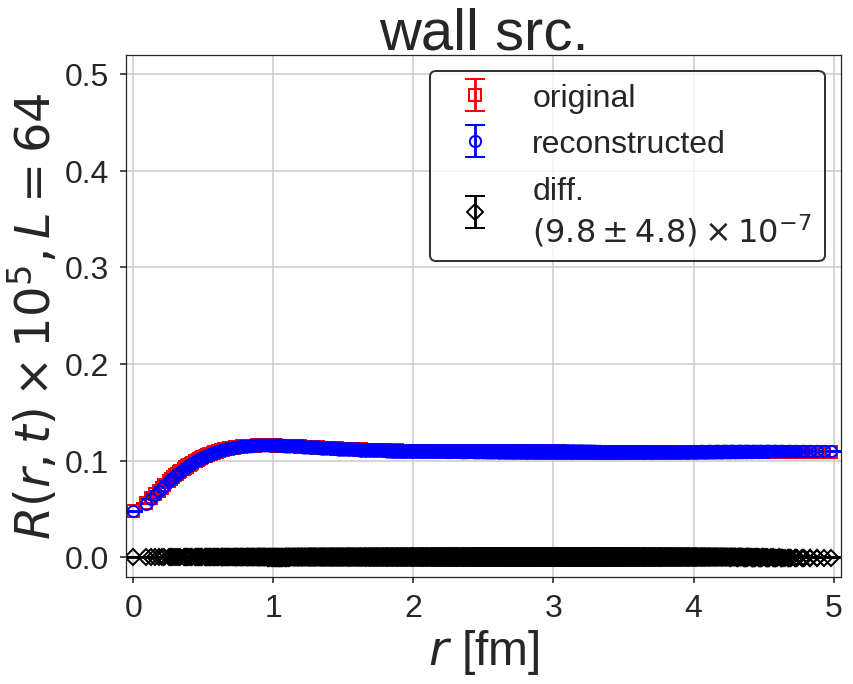}
  \includegraphics[width=0.48\textwidth,clip]{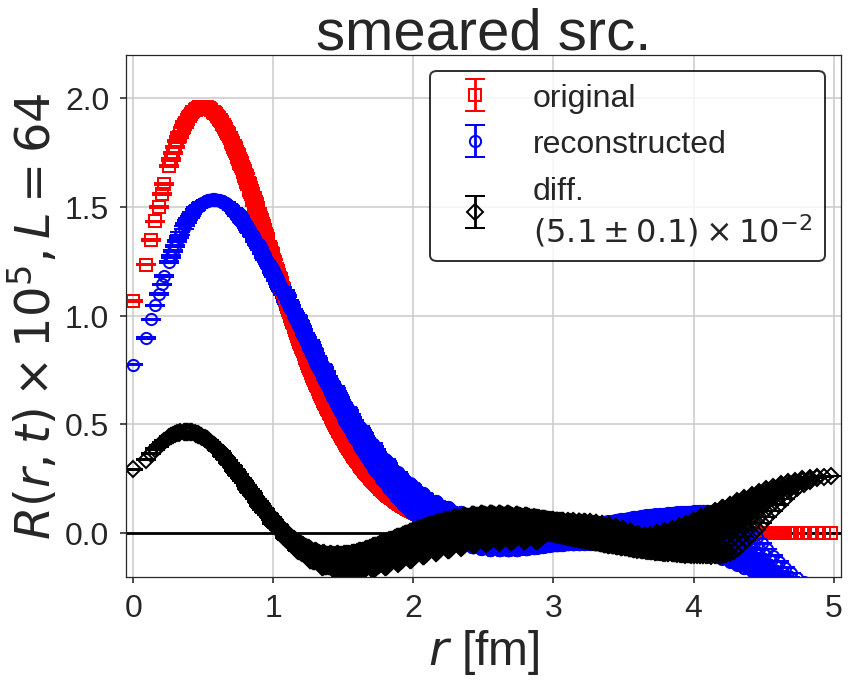}
  \caption{
    \label{fig:comp_ReR}
    Comparisons between the original $R$-correlator (red squares) and the reconstructed $R$-correlator 
    using low-lying eigenstates below the inelastic threshold (blue circles)
    at $t/a = 13$
    for $L=40$ (top), 48 (middle), 64 (bottom).
    The differences between the two $R$-correlators
      are shown by black diamonds and the values for the corresponding residual norms are given in legend panels.
    (Left) The wall source. (Right) The smeared source.
  }
\end{figure}

In this appendix, we study how good the original $R$-correlator $R(\vec{r},t)$ is approximated by
the reconstructed $R$-correlator $\overline{R}(\vec{r},t)$ given by
\begin{eqnarray}
  \overline{R}(\vec{r},t)
  =
  \sum_{n=0}^{n_\mathrm{max}} a_n \Psi_n(r) e^{-\Delta E_n t},
\end{eqnarray}
where $n_\mathrm{max} = 3$, 4, 6 for $L = 40$, 48, 64, 
corresponding to the number of excited elastic states below the inelastic threshold.
Fig.~\ref{fig:comp_ReR}
shows the comparison between
the original (red squares) and the reconstructed (blue circles)
$R$-correlators
for the wall source (Left) and the smeared source (Right) at $t/a=13$.
Shown together by black diamonds are the difference between the two, $R(\vec{r},t) - \overline{R}(\vec{r},t)$.
To estimate the magnitude of the difference,
we define the residual norm
by
\begin{equation}
  \frac{\sum_{\vec{r}}|R(\vec{r},t) - \overline{R}(\vec{r},t)|^2}%
       {\sum_{\vec{r}}|R(\vec{r},t)|^2},
  \label{eq:residual}
\end{equation}
whose values are also shown in the panels in Fig.~\ref{fig:comp_ReR}.

In the case of the wall source, it is found that the reconstruction works very well
and the residuals are negligible ($\lesssim 10^{-5}$).
In the case of the smeared source,
while the reconstruction is not as perfect as in the case of the wall source,
the reconstructed $R$-correlator well reproduces the overall behavior of the original one
where the residuals are only $\sim 0.4-5\%$.
The main origin of the residuals is
the small discrepancies at short distances
which become more apparent for a larger volume.
This indicates that there remain small contributions
from excited states above inelastic threshold
in the case of the smeared source.
In fact, since all elastic states couple to the
smeared source with the same order of magnitude
(See Fig.~\ref{fig:an_a0}),
there could remain contributions from elastic as well as inelastic states above the threshold for $R(\vec{r},t)$
even at $t/a=13$,
whose magnitude is expected to be larger for a larger volume due to the larger density of states.

\clearpage
\section{Reconstructed effective energy shifts}
\label{app:delta_E_r}

We collect the results on the reconstructed effective energy shifts,
Eq.~(\ref{eq:ReDEeff}). 

\subsection{The results on various volumes}
\label{subapp:delta_E_r:vol}

\begin{figure}[h]
  \centering
  \includegraphics[width=0.47\textwidth,clip]{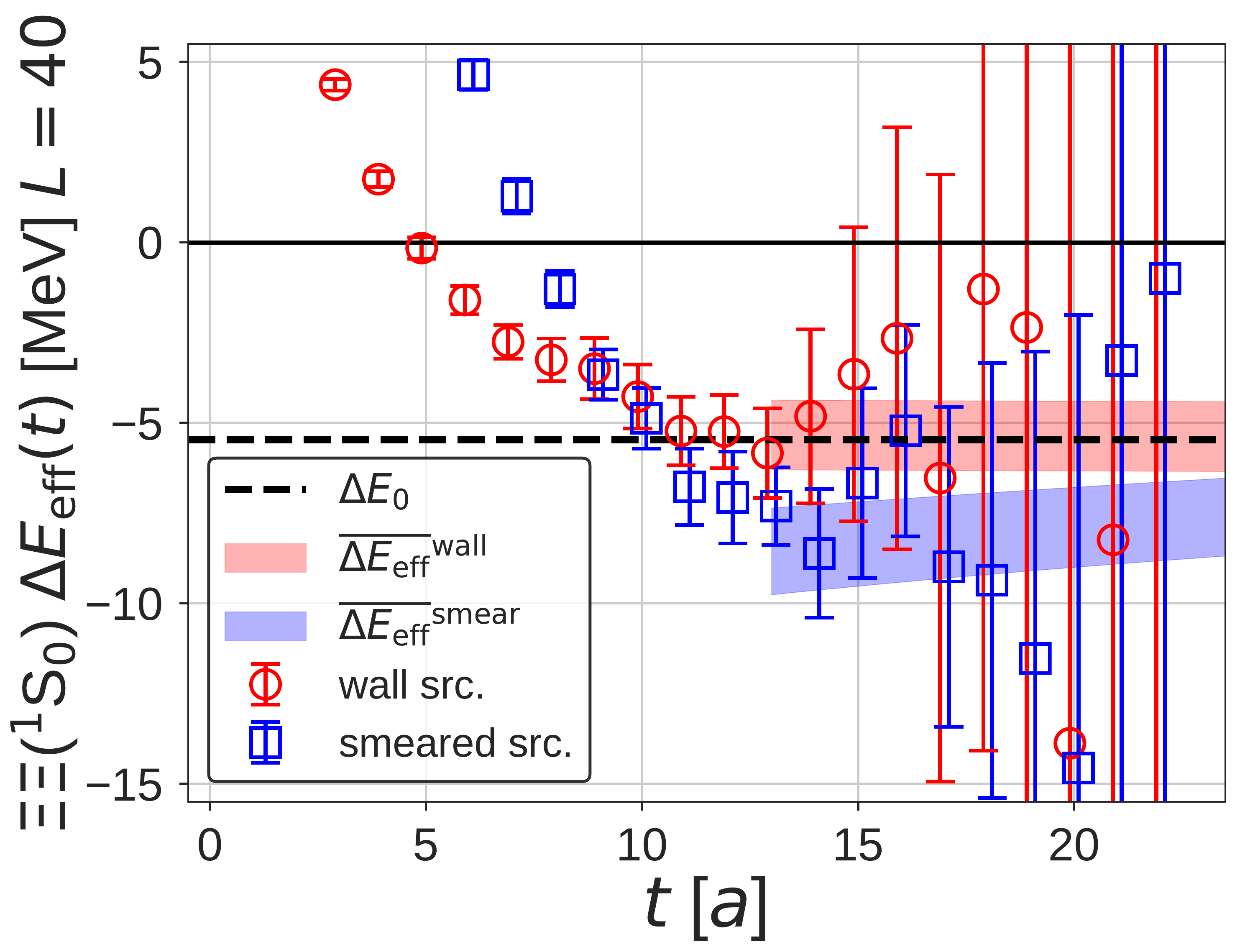}
  \includegraphics[width=0.47\textwidth,clip]{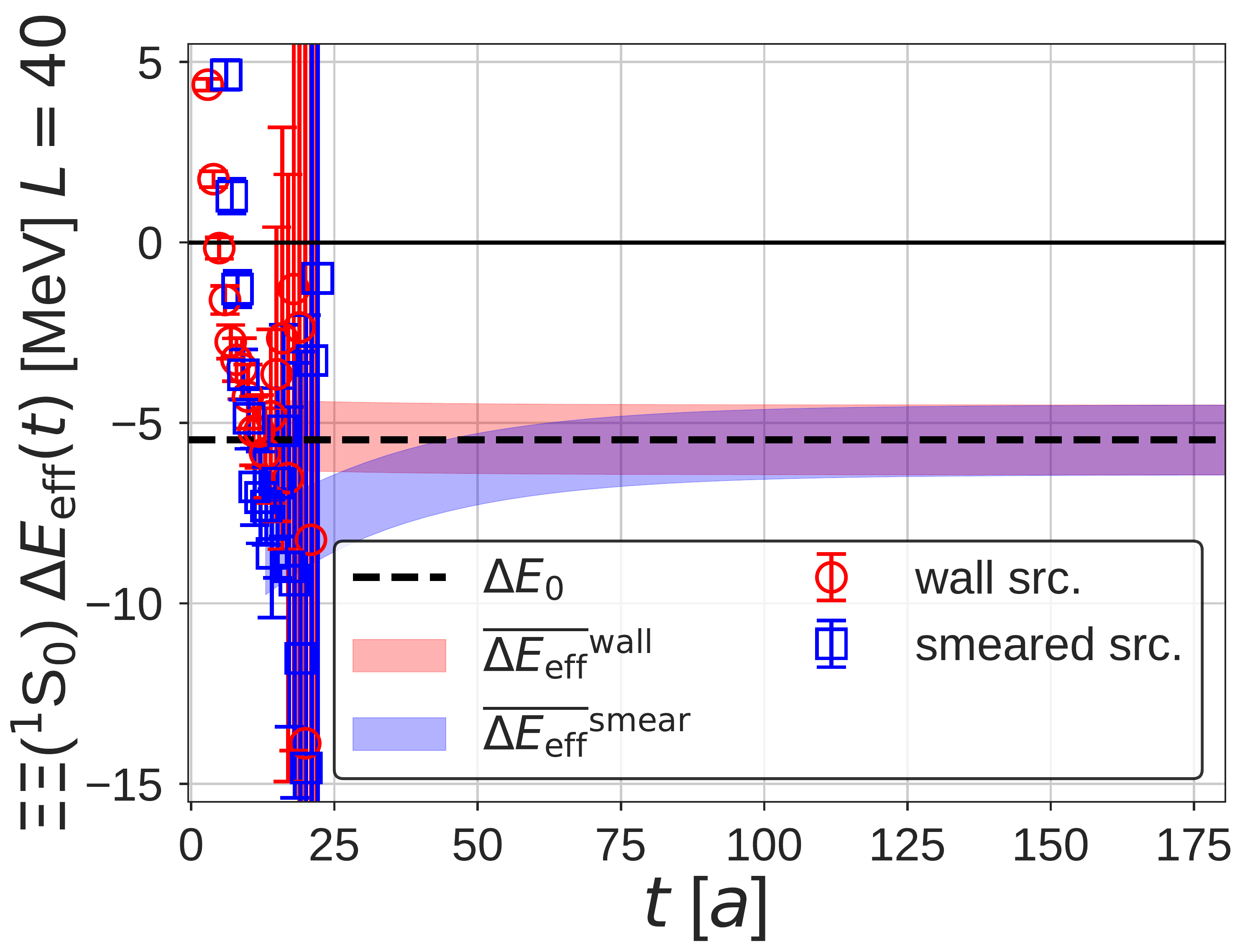}
  \includegraphics[width=0.47\textwidth,clip]{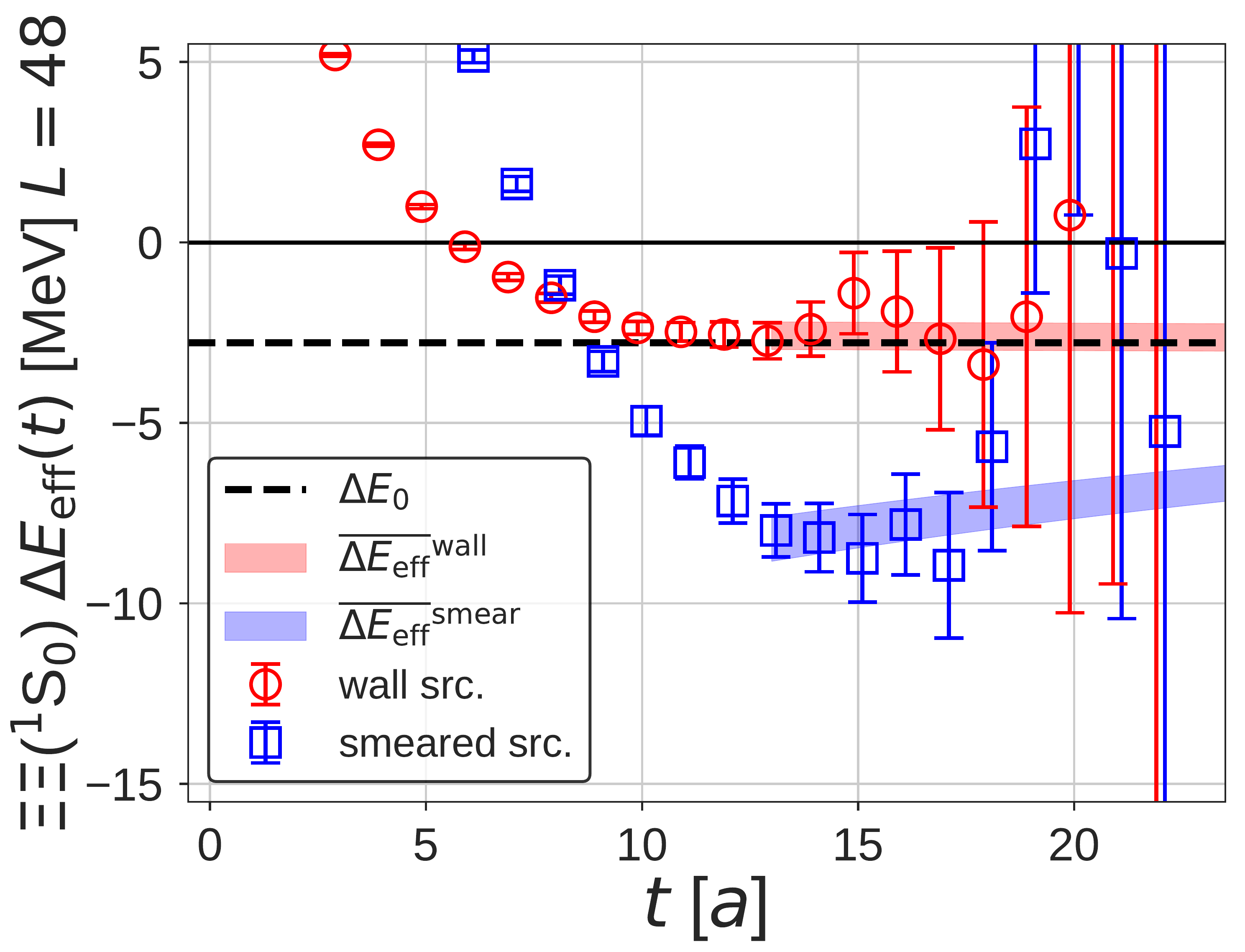}
  \includegraphics[width=0.47\textwidth,clip]{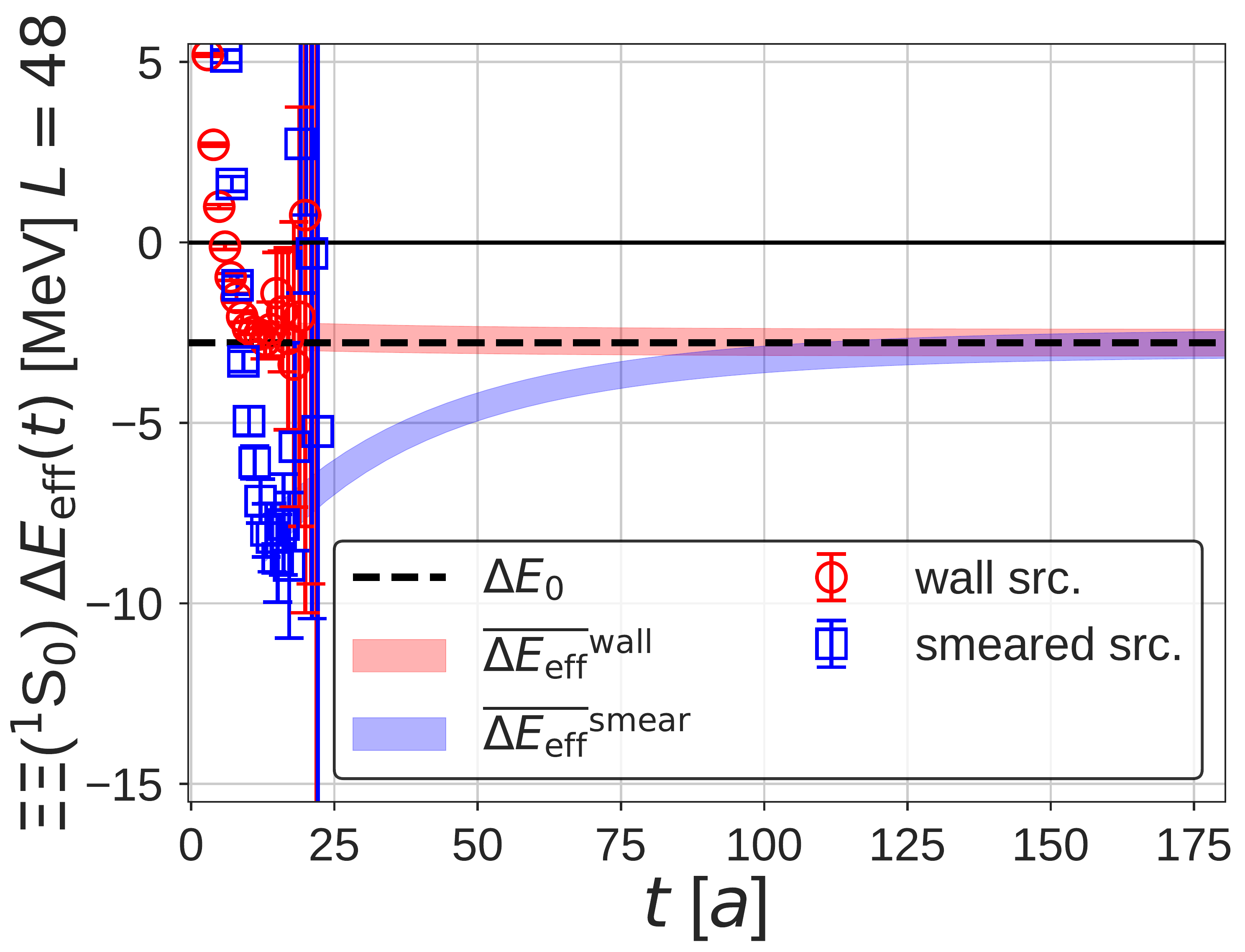}
  \includegraphics[width=0.47\textwidth,clip]{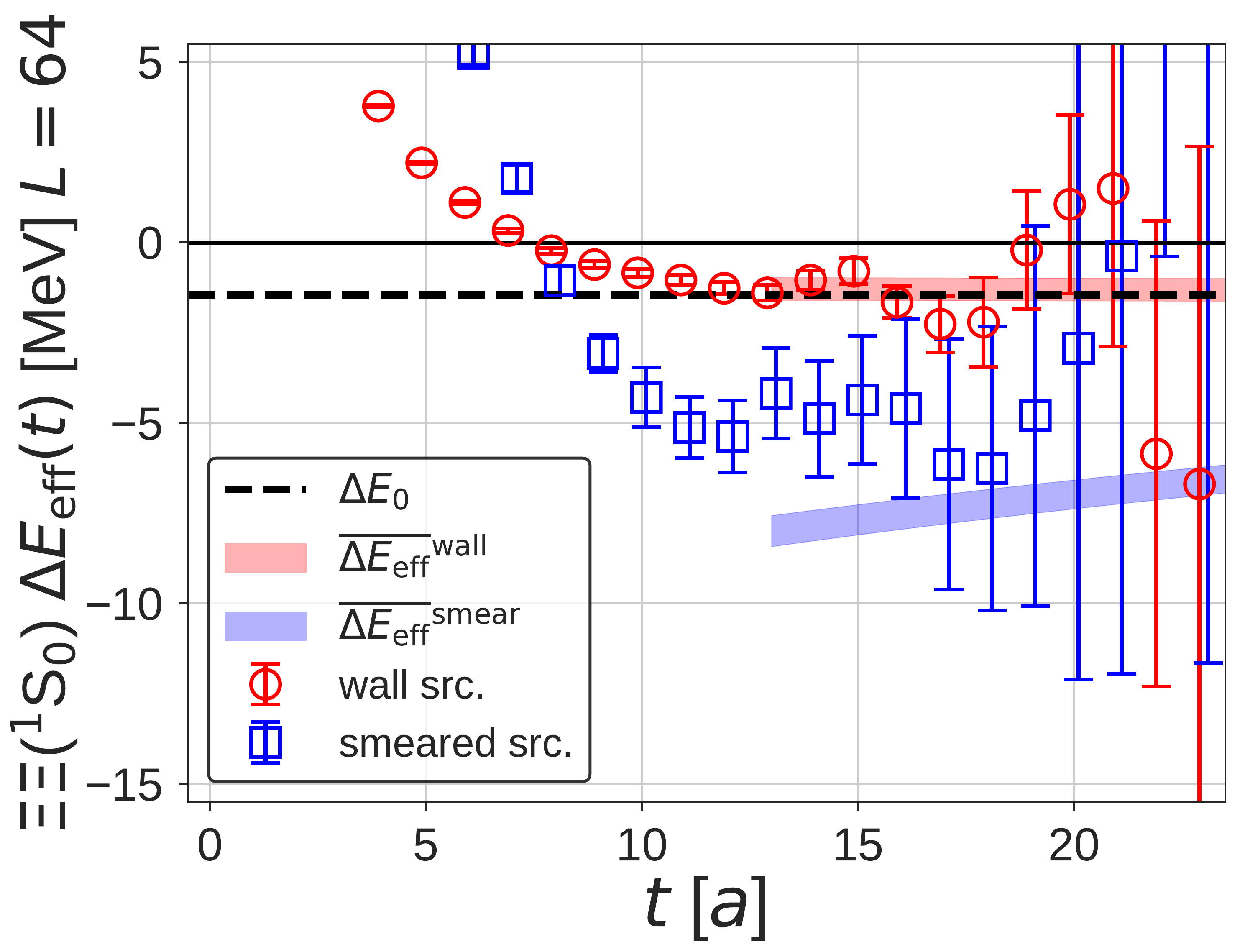}
  \includegraphics[width=0.47\textwidth,clip]{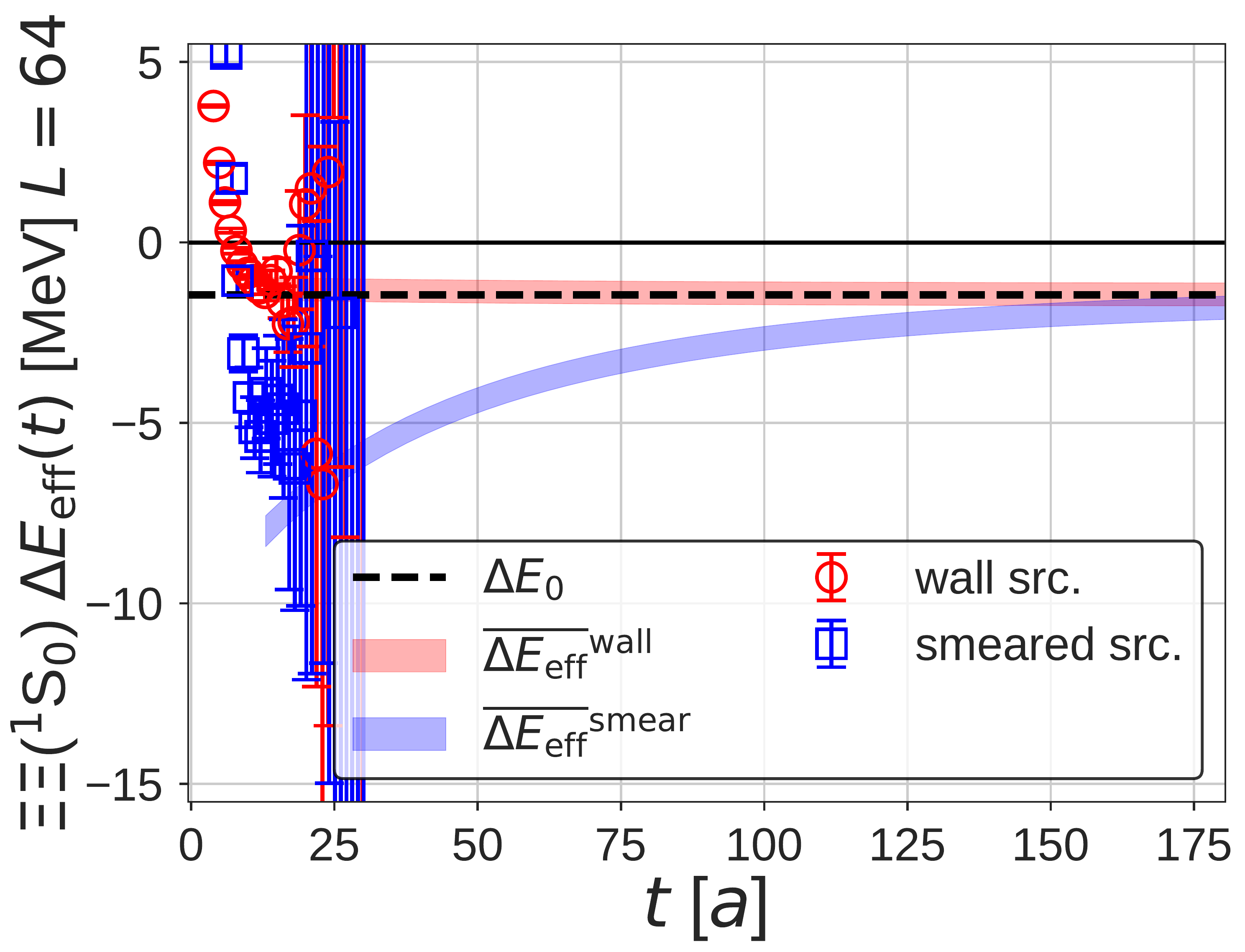}
  \caption{
    \label{fig:ReDEeffComp}
    The reconstructed effective energy shifts $\overline{\Delta E_\mathrm{eff}}(t, t_0 = 13a)$
    for the wall source (red bands) and the smeared source (blue bands) at $L = 40$, 48 and 64.
    The effective energy shifts in the direct method are also shown
    for the wall (red circles) and smeared (blue squares) sources.
    The black dashed lines are the energy shifts for the ground state of the HAL QCD Hamiltonian 
    in the finite volume evaluated at $t_0/a = 13$.
    (Left) $0 \le  t/a  \le 24$.
    (Right) $0  \le t/a  \le 175$.
}
\end{figure}

As shown in Fig.~\ref{fig:ReDEeffComp},
we observe that the reconstruction generally works well.
A small deviation at early time slices on $L=64$ for the smeared source
is most likely due to the statistical fluctuations and/or
the systematics due to contaminations from the states above the threshold
 discussed in Appendix~\ref{app:inelastic}.

\clearpage
\subsection{Contributions from excited states to the effective energy shifts}
\label{subapp:cut_off}

We study how each elastic excited state
contributes to the effective energy shift
for the smeared source,
by changing the number of elastic excited states 
 used in the reconstruction ($n_\mathrm{max}$)
in Eq.~(\ref{eq:ReDEeff}).

\begin{figure}
  \centering
  \includegraphics[width=0.5\textwidth,clip]{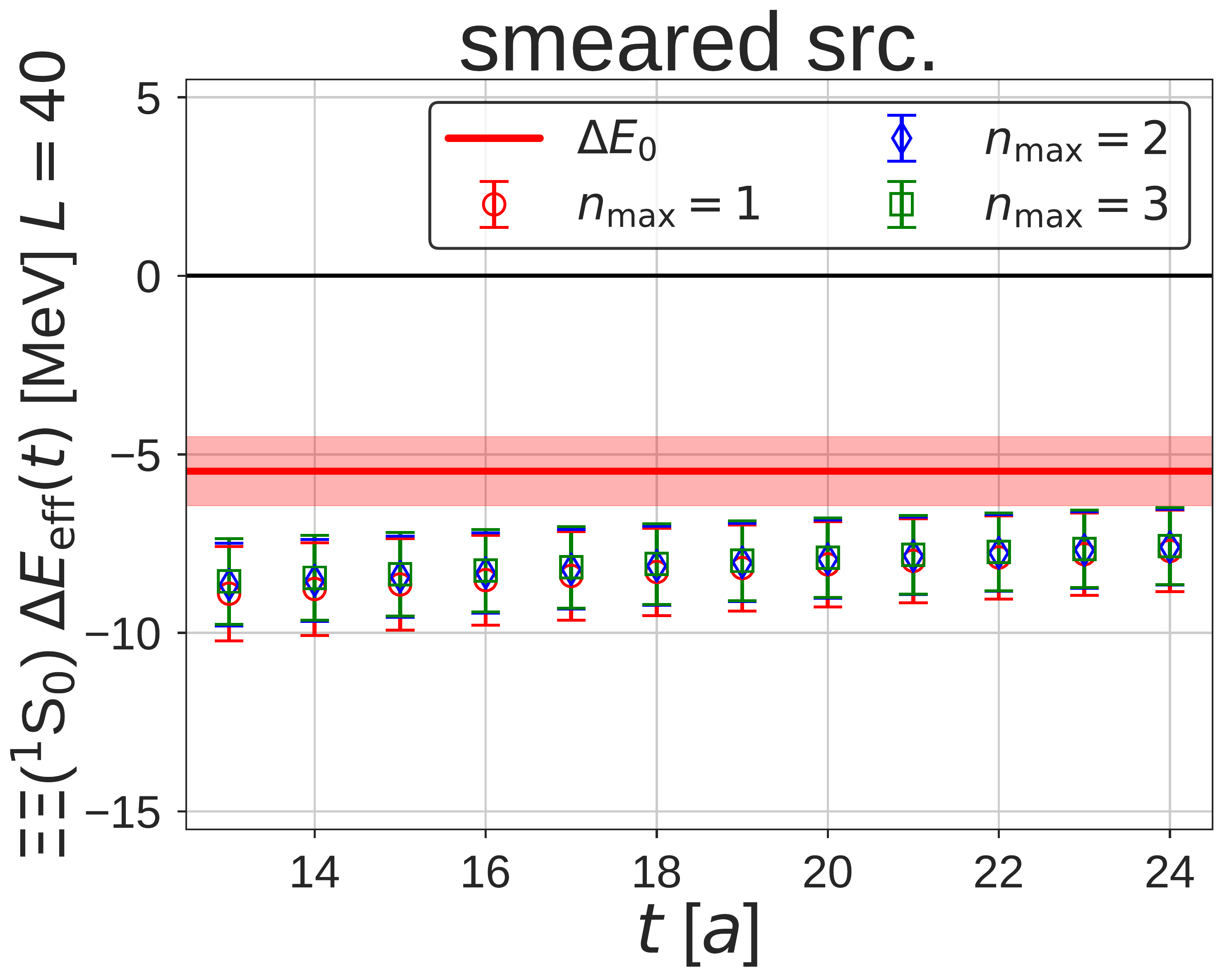}
  \includegraphics[width=0.5\textwidth,clip]{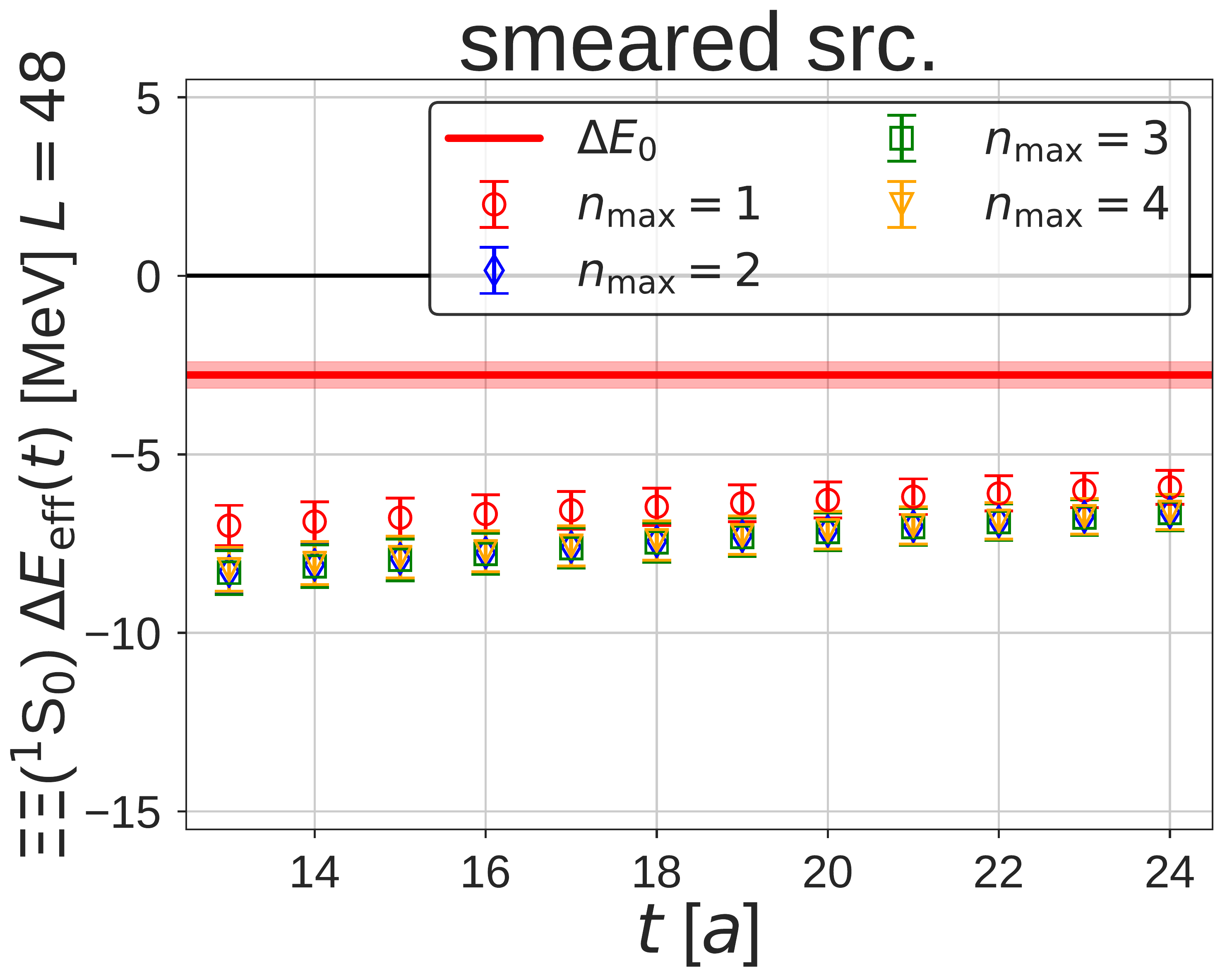}
  \includegraphics[width=0.5\textwidth,clip]{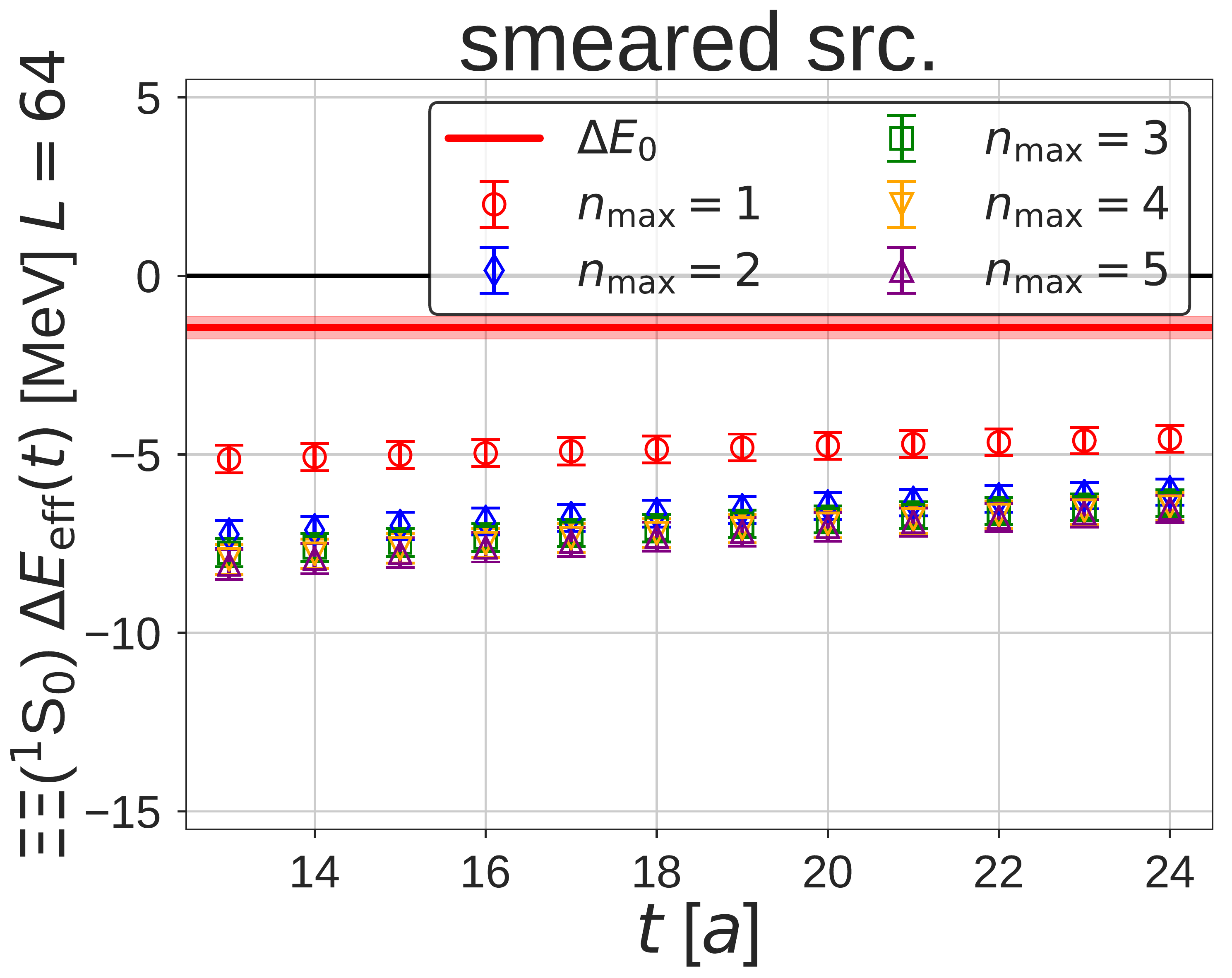}
  \caption{
    \label{fig:ReDEeff_n_dep}
    The mode number ($n_\mathrm{max}$) dependence of the reconstructed effective energy shift
    $\overline{\Delta E_\mathrm{eff}}(t, t_0 = 13a)$
    for the smeared source at $L = 40$ (top), $48$ (middle) and $64$ (bottom).
  }
\end{figure}

Fig.~\ref{fig:ReDEeff_n_dep} shows the $n_\mathrm{max}$ dependence of 
the reconstructed effective energy shifts for the smeared source
at $L = 40$, $48$ and $64$,
which are compared with the lowest eigenenergies $\Delta E_0$ (red bands).
Due to the negative sign of $b_n/b_0 < 0$ (see Fig.~\ref{fig:bn_b0}~(Right)),
the reconstructed effective energy shifts are smaller than $\Delta E_0$.
For $L = 40$ and 48,
the dominant contribution  comes from the first excited state ($n_\mathrm{max} = 1$) besides the ground state
and higher modes give only minor corrections,
while the second excited state ($n_\mathrm{max} = 2$) also gives  significant contribution
for $L=64$.
These results indicate that the pseudo-plateau structures around $t/a \sim 15$
are originated mostly from the scattering states below $\sim 90$ MeV (See Table~\ref{tab:deltaEn_summary}). 

\clearpage
\section{Effective energy shifts from the improved sink operator based on eigenfunctions}
\label{app:eigen-proj}

In this appendix, we present  effective energy shifts on various volumes,
which are obtained from the improved two-baryon sink operator,
 projected to the ground state/first excited state in Eq.~(\ref{eq:EeffProj}).
 Results 
are found to be consistent with the corresponding finite-volume eigenenergies of the HAL QCD Hamiltonian.
Small discrepancies for the smeared source on $L=64$
are most likely due to the statistical fluctuations and/or
the systematics 
 due to contaminations from the states above the threshold
 discussed in Appendix~\ref{app:inelastic}.
Shown together in the left figure are the effective energy shifts in the direct method (without projection),
where the significant deviation is observed for the smeared source.

\begin{figure}[h]
  \centering
  \includegraphics[width=0.47\textwidth,clip]{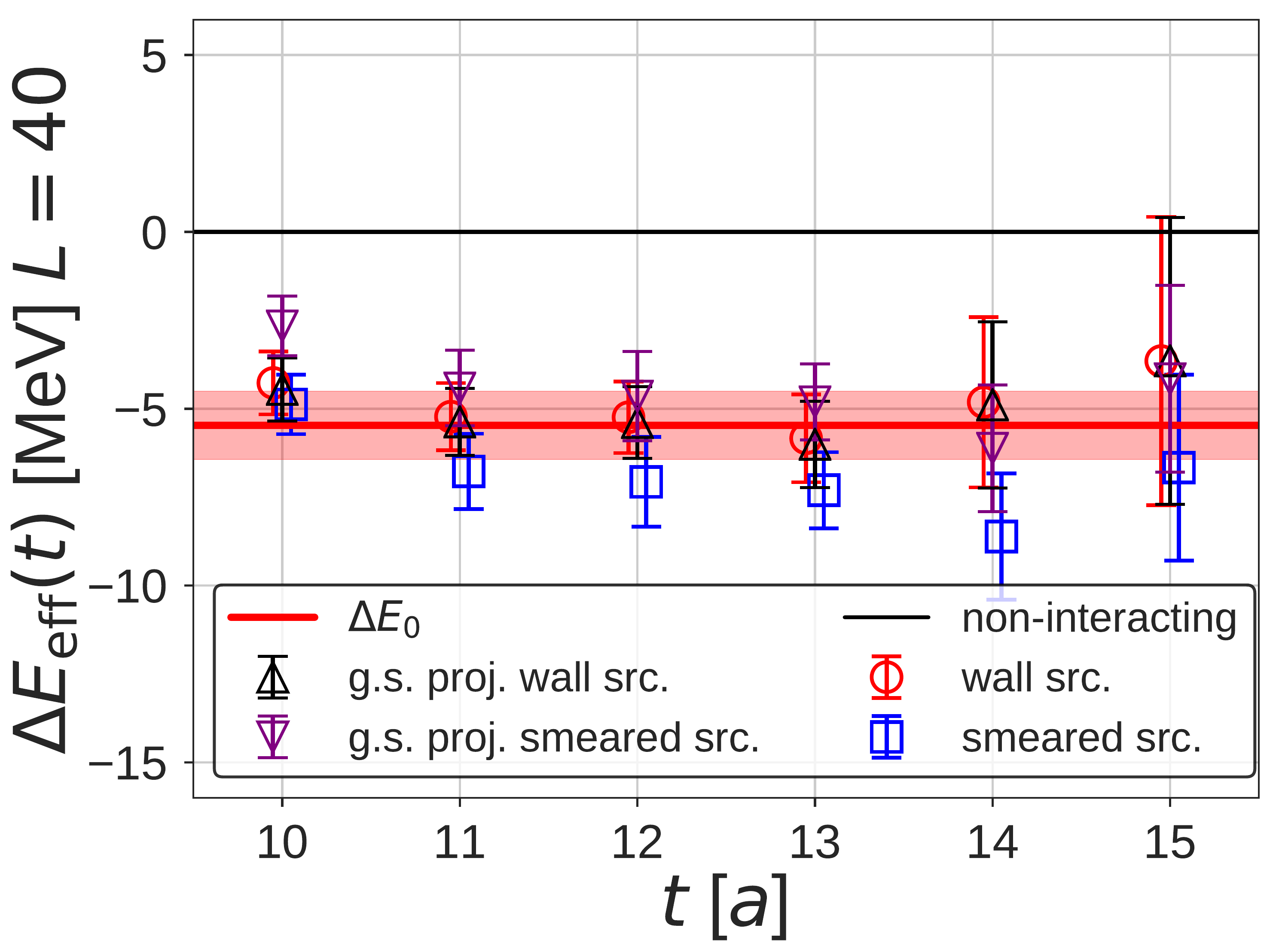}
  \includegraphics[width=0.47\textwidth,clip]{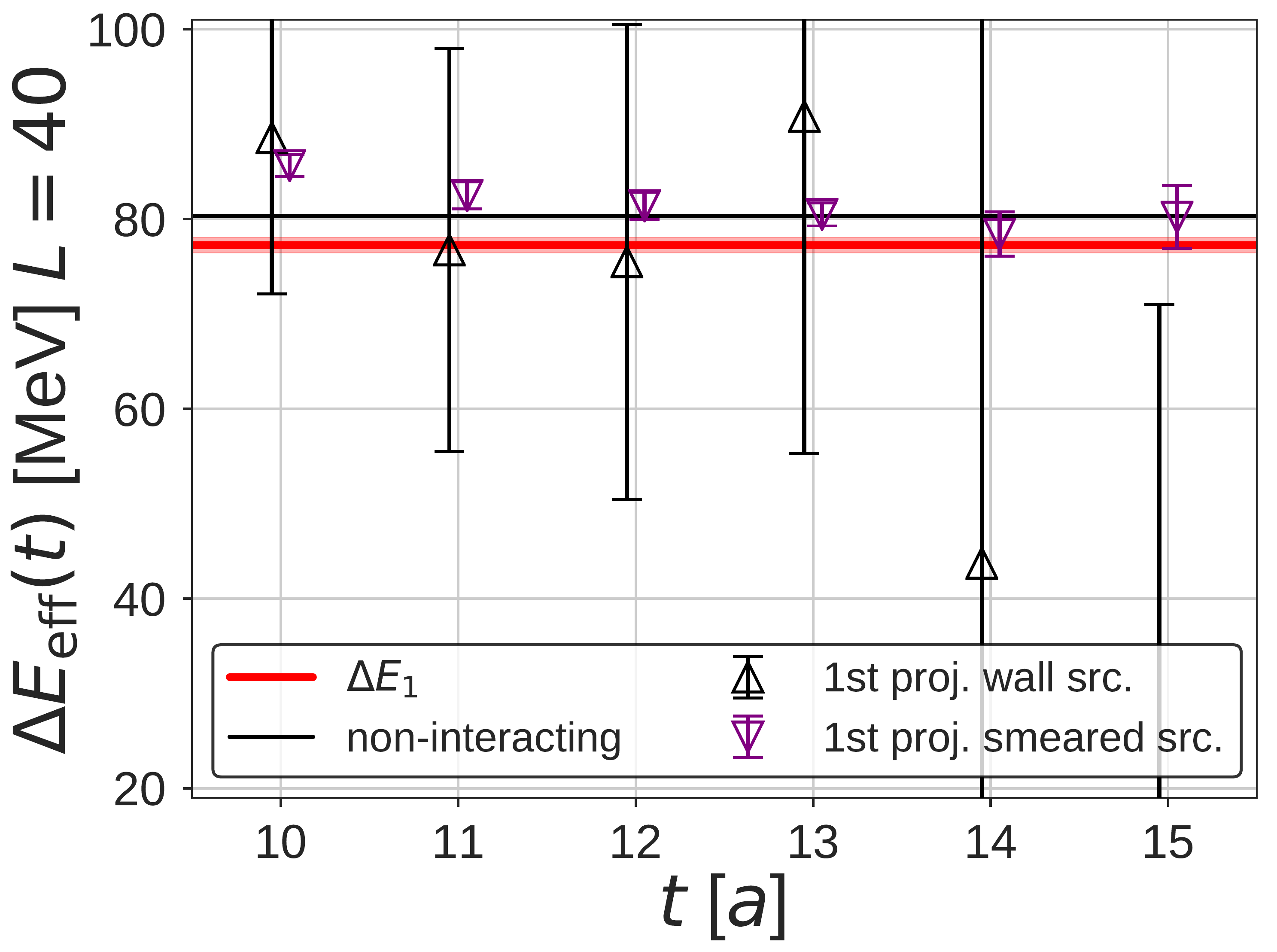}

  \includegraphics[width=0.47\textwidth,clip]{figs/dEeffs/gs_projected_L48.pdf}
  \includegraphics[width=0.47\textwidth,clip]{figs/dEeffs/1st_projected_L48.pdf}

  \includegraphics[width=0.47\textwidth,clip]{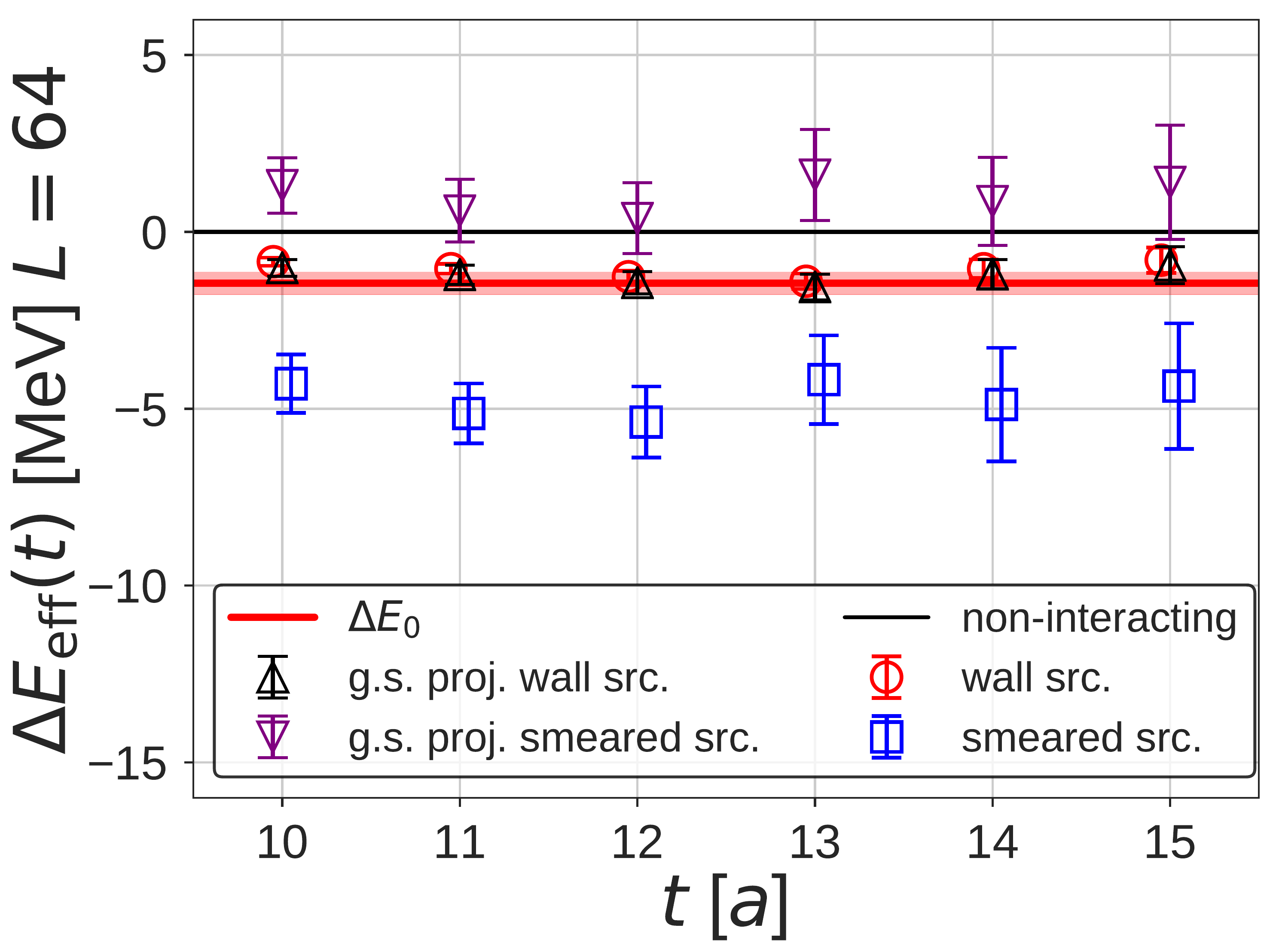}
  \includegraphics[width=0.47\textwidth,clip]{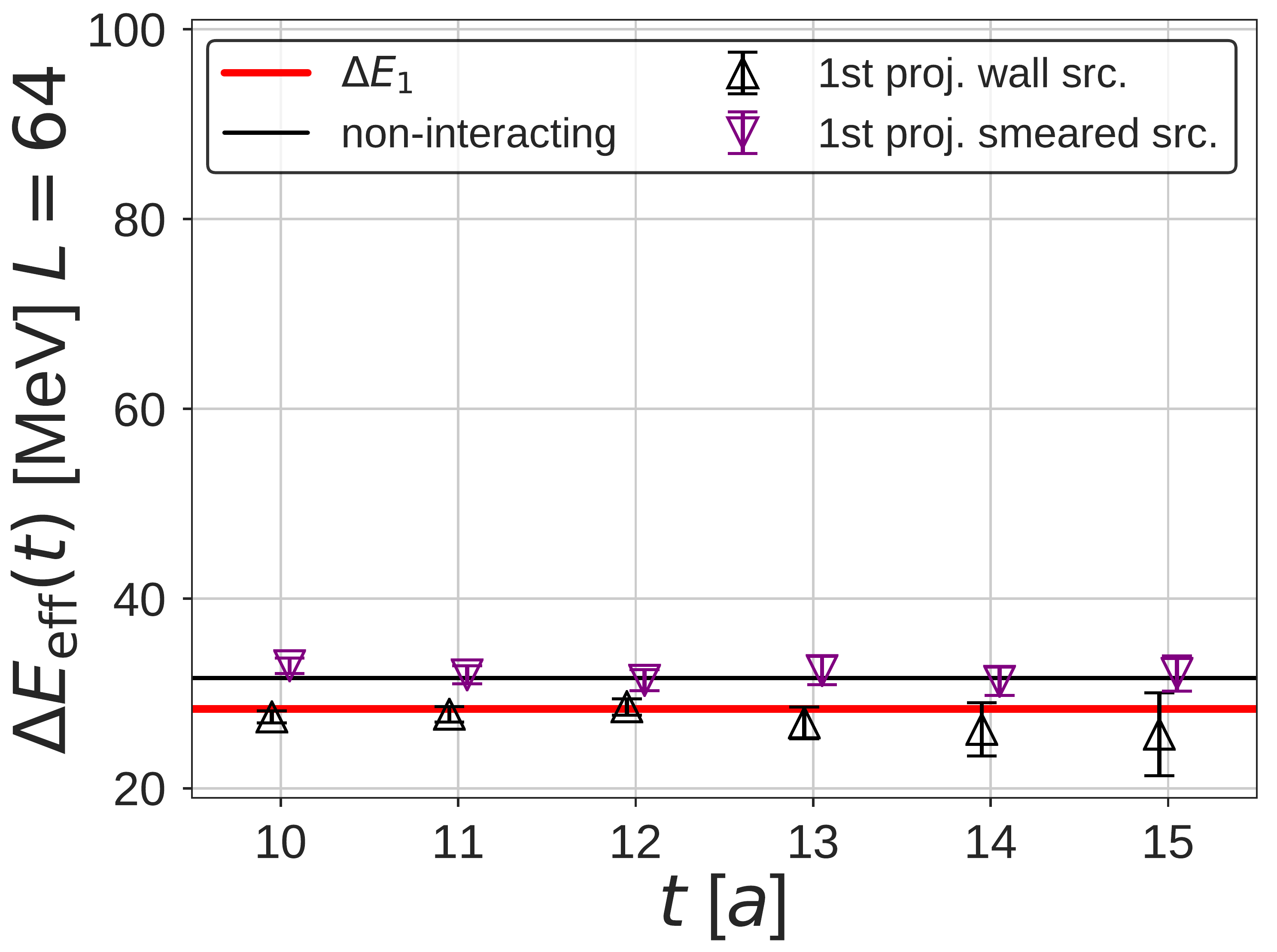}
  \caption{
    \label{fig:Eeff_proj}
    Same as Fig.~\ref{fig:Eeff_proj:48}
    on $L = 40$ (top), $48$ (middle) and 64 (bottom).
  }
\end{figure}

\end{document}